\documentclass[preprint2]{aastex}

\slugcomment{}
\shorttitle{Chandra Observations of Early-type Galaxies}
\shortauthors{Athey}

\begin{document}

\title{The Origin and Evolution of the Interstellar Medium in Early-type Galaxies \\ Chapter IV -
Chandra Observations of Early-type Galaxies}

\author{Alex E. Athey \altaffilmark{1}}
\affil{University of Michigan }
\affil{Department of Astronomy}
\affil{500 Church St.}
\affil{Ann Arbor, MI 48109-1090}
\affil{December, 2003}
\altaffiltext{1}{current email athey@arlut.utexas.edu}

\author{Ph.D. Advisor: Joel N. Bregman}
 
\begin{abstract}
We have performed uniform analysis of a sample of 54 nearby, early-type galaxies observed with \emph{Chandra}.  
In this work we present the spectral results for both the diffuse 
Interstellar Medium, ISM, and low-mass X-ray binaries, LMXBs.  We determine 
the metallicity of the hot ISM in the 22 brightest galaxies and find a narrow range of abundance
ratios relative to iron. The average iron metallicity of these bright galaxies is $0.96\pm0.33$ 
relative to solar. 
By assuming these enrichment patterns continue to fainter galaxies,
we are able to
extend accurate ISM modeling down to the faintest galaxies in the sample.  
The sample of galaxies span 4.5 orders of magnitude in
X-ray luminosity starting at $L_X=10^{38}\, erg \cdot s^{-1}$ and the average gas temperature in 
the sample is
$0.58\pm0.24\, keV$.  We present the X-ray 
properties of these galaxies scaled to one effective radius as well as radial variations 
of gas and stellar binary luminosities, and radial variations of gas temperature, 
metallicity, mass, entropy and density.  
\end{abstract}

\keywords{X-rays, Early-type Galaxies, Elliptical Galaxies, Chandra, Hot ISM, LMXB, Hot ISM Metallicity}

\section{Introduction}

Ever since the discovery that most early-type galaxies host hot, X-ray
emitting gas halos \citep{Forman_Einstein}, attempts have been made
to understand the origin and evolutionary fate of this gas
and how this relates to the formation history of the galaxy itself.
This gas provides
a probe of the gravitational potential, and signatures of recent tidal 
or cluster interactions can be detected by examining the mass and
temperature of the gas.
Although great progress has been made, after \emph{Einstein},
\emph{ROSAT}, and \emph{ASCA}, and the first years of \emph{Chandra}
and \emph{XMM}, questions remain and the discord between the properties
of the interstellar medium (ISM) detected in X-rays and the predicted
characteristics from other waveband data has not been adequately resolved. For
example, the metallicity of the ISM is expected to be at least solar
since the stars shedding their envelopes into the ISM have optically
measured metallicities near solar (e.g. \citet{Trager_2,Trager_1,Forbes_Ages}),
but \emph{ASCA} data were unable to discriminate between extremely
sub-solar metallicity \citep{ASCA_NGC4636_Mushotzky}, solar metallicity
\citep{ASCA_NGC4636} or even two plasmas with different temperatures
and higher metallicities \citep{Buote_2-Temps_1,Buote_2-Temps_2}.
Irrefutable metallicity determinations from \emph{XMM's} Reflection Grating Spectrometer
(RGS) have shown the metallicity of several, nearby and bright galaxies
to be near solar metallicity, and with non-solar metal-to-metal ratios
\citep{XMMRGS_NGC4636,XMMRGS_NGC5044}, but these results have not
yet been reproduced from the \emph{Chandra} imaging data, which does a better
job of isolating the diffuse gas than \emph{XMM's} slitless RGS. It is also
not clear if the two galaxies examined are representative of all nearby early-type
galaxies.

The greatest disparity between the inferred properties of the gas from
 X-ray observations
and other waveband observables concerns cooling flows. While there is
some debate about the metallicity of the gas, the density of the gas
is well determined, typically  $\sim10^{-2}n_{e}cm^{-3}$ in the center of
an early-type galaxy. It is a straightforward observation
to determine the density as the emission measure varies as $n_{e}^{2}$,
although, this implies a cooling time of $10^{7}yrs$ which is at odds
with the observed equilibrium of the gas and a wealth of other waveband
observations (e.g. the lack of star formation seen in normal early-type
galaxies). Recently, the \emph{FUSE} observatory has been used to
detect cooling gas \citep{Joel_OVI_FUSE}, but at much lower rates
than implied by the densities observed in the X-rays. Many of these galaxies
have been surveyed for {[}O II{]} emission (e.g. \citet{Caon_paper_I})
and FIR emission (e.g. \citet{Goudfrooij_2}) in efforts attempting
to detect hidden star formation, but there are no strong detections.
This non-detection of cooling likely means an additional heating source,
but all the likely candidates (SNe and/or AGN) have inconsistent
rates, energetics, spatial, or spectral distributions from the 
needed missing heat source.

Observations of individual galaxies provide important contributions
to the field, especially when new satellites are launched, however,
the baseline of knowledge of the hot gas in early-type galaxies is derived
from large (sample size $\gtrsim$ 25) studies 
(See Table \ref{Large Surveys} for large X-ray studies of early-type
galaxies). In fact, it was through the large studies that the existence
of the hot gas was first discovered as a universal phenomenon in early-type
galaxies \citep{Einstein_2}. These large X-ray studies seek to address one or more
of the following goals: 1) infer the physical source(s) of the X-ray
emission, 2) calculate accurate global plasma temperatures, densities
and metallicities and measure radial variations of these quantities,
3) characterize the morphology of the hot gas via surface brightness
profiles, 4) determine the scaling relations between global galaxy
properties: X-ray luminosity, temperature, stellar velocity dispersion
and optical stellar luminosity, 5) detect any influences the environment
may have on the X-ray properties. Depending on the technical capabilities
of the X-ray mission employed, the large surveys had differing levels
of success in addressing each of these goals.  \citet{Trevor_ROSAT_study}
notes that an order of magnitude disparity exists in the methods for 
statistically removing
the point source contribution and isolating the diffuse emission from
the low-mass, X-ray binaries, LMXBs. The spatial resolution of \emph{Chandra} allows for the
isolation of the point sources from the diffuse emission which has 
been the major source of uncertainty in previous studies of the hot ISM in fainter
galaxies.

A large study with the \emph{Chandra} satellite provides significant
advances in addressing the large study goals due to its superb spatial
resolution and moderate spectral resolution. As of 2003, no large study
of nearby, early-type galaxies have been published with \emph{Chandra} or
\emph{XMM-Newton} observatories despite the maturity of the archives.
Although \emph{XMM} has a larger collecting area, it is the fine spatial
resolution of \emph{Chandra} that will provide the next critical step
in understanding X-ray emission from early-type galaxy as the majority of the ``containmenating''
stellar binary component can finally be removed from gas studies. 

Motivated by the scientific returns of a large study
of early-type galaxies with \emph{Chandra} and the availability of
over five dozen galaxies in the public archive, we conducted an extensive
program investigating all components of the X-ray emission in early-type
galaxies. In Section \ref{Chandra Sample} the sample is defined and
limitations of a \emph{Chandra} archival study are discussed. In Section \ref{Chandra Data}
the data reduction process is described, including efforts taken to
correct the known calibration issues. Section \ref{Analysis} discusses
the spectral analysis method and the success of its application to
the data. Section \ref{Chandra Results Spectral} presents the results of the
spectral fitting of the data and, where sufficient counts are available,
radial variations of the physical parameters fitted in the models.
In Section \ref{Chandra Results Cumulative} global X-ray properties
of the sample are presented. 
We compared the luminosities and temperatures derived with \emph{Chandra} to
\emph{ROSAT} in Section \ref{ROSAT Comparison}.

~

~

\section{\label{Chandra Sample}Galaxy Sample}

The sample selection criteria is simply defined as any nearby early-type
galaxy observed by \emph{Chandra} and in the archive.
Since our methodology depends on
the low-energy response of the backside illuminated chips of the ACIS
array, we restrict our study to ACIS-S3 observations. In Table
\ref{Chandra Sample Table} the 54 galaxies in the archive that meet
this selection criteria along with their distances, blue magnitudes,
X-ray fluxes determined with \emph{ROSAT} and effective radii are
listed. All the galaxies are relatively nearby, with distances between
10 and 110 Mpc, and bright, with $m_{B}^{0}<13$, and most have been 
studied in one of the previous large studies (Table \ref{Large Surveys}).
Also listed in Table
\ref{Chandra Sample Table} is any peculiar property of the galaxy;
included in the sample are a number of dwarf, starburst, merging, 
isolated, interacting and brightest
group galaxies which, for various reasons, may fall outside the realm
of normal early-type galaxy characterization.


The relatively small field of view of the \emph{Chandra} ACIS-S3 chip
($\sim8\arcmin$), presents a limitation to accurately determining
global properties for large nearby galaxies which have halos detected
with \emph{ROSAT} out to tens of arcminutes. But since the X-ray emission
is sharply peaked, this correction is small and if needed, can be
extrapolated from \emph{ROSAT} or \emph{XMM} data. We note that the
three of the four brightest galaxies are also the most distant, preserving a general
on-chip size scale for the sample as a whole (i.e. even the largest galaxies can be observed
out to one effective radius without vignetting).

\section{\label{Chandra Data}Data Reduction}

The data were processed in a uniform manner following the \emph{Chandra}
{}``scientific threads''%
\footnote{http://cxc.harvard.edu/ciao/threads%
} employing CIAO 3.2 coupled with CALDB 3.0. All of the data were
reprocessed with the most recent standard gain maps provided. 
The 5x5 array telemetered for each event in the {}``Very
Faint'' observing mode 
allowed for improved screening of cosmic rays
and decreases the background by about a factor of 1.25 for S3 observations.%
\footnote{http://cxc.harvard.edu/cal/Links/Acis/acis/Cal\_prods/vfbkgrnd/index.html%
} Otherwise the standard cosmic ray rejection method was used. ASCA
grades of 0,2,3,4 and 6 were selected for all subsequent processing
and analysis. Bad pixel maps, specific for each observation, were
applied from the standard calibration library.
In galaxies with strong AGN activity, pile-up becomes a concern, so the
central regions in these galaxies are excluded from the diffuse and point
source analysis.
The data were corrected for the charge-transfer inefficiency that
degrades the spectral resolution of the imaging devices 
with the standard processing techniques for the -120 temperature data.

We checked for background particle flares via a temporal background rate
light-curve from the S1 chip if the S1 chip was read out with the
S3 chip following the data cleaning procedures described in \citet{Markevitch}.
In about half of the observations, the S1 chip was not read out (only
six of the ten ACIS chips can be read out for any given observation
and these six are chosen by the observer).
For these observations
the temporal light-curve was constructed
from the outer regions of the S3 chip. 
The energy sampling was restricted
from 2.5-7.0 keV to increase sensitivity to the most common type of
flares detected \citep{Markevitch}. For the temporal filtering, a
mean background rate was determined by isolating a quiescent period
and a 2.5$\sigma$ deviation was determined from this non-active period.
This quiescently defined mean and clipping sigma were applied to the
data.
The aggressive, 2.5$\sigma$
filtering was applied since the soft counts included from the flares
can influence the spectral fitting of low count, spatially extended,
soft energy emission of the diffuse gas. When working with the point sources
a 3.5$\sigma$ rejection was applied since the hard and spatially
isolated spectrum of the point sources is less affected by the flares.

We applied a correction to adjust for the gain changes related to
charge transfer inefficiencies that produce a drift in the PI channels,
or effectively a drift in energy or wavelength. Starting with CALDB 2.26,
the spectral response-table also corrects for the low-energy absorption,
thought to be occurring from the out-gassing of unknown materials in
the spacecraft and the subsequent condensation on the blocking filters
(e.g. \citet{Chandra_Absorption}). Previously, this low energy degradation
was corrected with the ACISABS model%
\footnote{http://cxc.harvard.edu/cont\-soft/software/ACISABS.1.1.html%
} and our tests confirm that both these corrections produce similar
and acceptable results for low energy work.

In order to have an independent gauge of the background, the calibration
team's blank sky observations (acis7sD1999-09-16bkgrndN0001.fits for
-110 data and \newline acis7sD2000-12-01bkgrndN0002.fits for -120 data) were
employed. For greater consistency, these observations were reprocessed
with calibrations consistent with the processing described above.
In each of the galaxy observations, the peak energy distribution
of the diffuse gas was determined and used to generate a mono-energetic
exposure map.

Sources were detected using the {}``Mexican-Hat'' wavelet detection
routine \emph{Wavdetect} in a 0.3-6.0 keV bandpass, chosen to minimize
the background. The threshold was set to give approximately one false
detection per image and PSF scales were run from 
$\sqrt{2}^{0}-\sqrt{2}^{8}$ in 8 steps, 
allowing for a range of extended sources to be detected.
All of the sources were extracted from the image with annular background
regions of a few pixels around each source, creating a local background
and eliminating any underlying diffuse count contamination. Count
weighted response matrices were calculated for the sources as a group.
The combined source spectrum was adaptively rebinned to contain a minimum of
25 counts per bin for reduced statistical errors in subsequent analysis.
Both the central source in the galaxies, frequently an AGN, and the
likely background AGN ($L_{X}>2\cdot10^{39}erg\, s^{-1}$) \citep*{Jimmy_ULX},
were separated from the rest of the sources, leaving a spectrum dominated
by LMXBs.

\section{\label{Analysis}Spectral Analysis}

The spectral analysis is performed with the goal of obtaining
maximum comparative information between the galaxies. The challenge
is to extract the maximum amount of information from the photon-rich,
bright galaxies where total counts can be as high as $\sim100,000$ in an exposure
while at the same time obtaining comparative results
from the photon-starved, faint galaxies where the diffuse gas counts
can be less than 500. The developed analysis procedure constructs
a framework of common methods where the quality of the data determine
both the amount of information extracted and the aggressiveness of the
removal of higher-order undesired signal. In this manner we create
a sample as uniform as possible given the large dynamic range presented
in the data.

\subsection{\label{Chandra Backgrounds and Regions}Backgrounds and Region Definitions}

With the sources identified, the diffuse counts can be isolated
and extracted from the image into annular bins. Three different background
determinations were employed: 1) An outer region on the chip where
the diffuse counts visually appeared to fall into the background,
2) Equal region descriptions in the source chip and the blank sky
observations, and 3) A local background subtraction, described below.

\begin{figure}
\epsscale{0.5}
\plotone{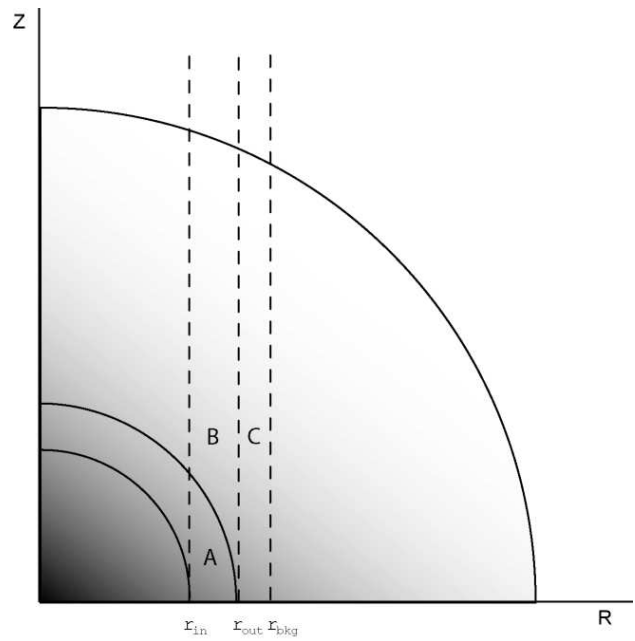}
\caption[Schematic picture of the local background subtraction method]
{Schematic picture of the local background subtraction method.
The observer is located at a large distance along the Z-axis
and spherical symmetry is assumed for galaxy emission. The region of
interest is from $r_{in}$to $r_{out}$, or area A, the union of the
spherical annulus of interest with the observational column probed.
However, the observation yields a column of emission which is equal to area
A + B. Given a smooth distribution of galaxy emission, the column
of emission represented by area B can be approximated by an area just
outside of $r_{out}$. The background radius, $r_{bkg}$, is chosen
so that the emission within a volume is the same for areas B and 
C.}
\label{BackgroundFig}
\end{figure}

The local background subtraction method is designed to be a simple
background subtraction that attempts to recover more of the true 3-dimensional
radial properties than the traditional 2-dimensional column probing,
while avoiding some of the complications of 3-dimensional deprojection
methods. A schematic picture of this method is pictured in Figure
\ref{BackgroundFig}. In an annular column probe through a galaxy, emission is seen from
both the thin annular shell of interest and emission from regions
outside of that shell. If the emission is spherically symmetric and
changes predictably with radius, it is possible to subtract an adjacent
column of emission that approximates the excess emission observed
in the source region.

In practice, we fit a $\beta$-model to the surface brightness profile, 
SB, of the galaxy,
\begin{equation}
SB\propto(1+(\frac{r}{r_{c}})^{2})^{-3\beta+\frac{1}{2}}
\end{equation}
and use this determination for a description of the emission measure,
EM, 
\begin{equation}
EM\propto(1+(\frac{r}{r_{c}})^{2})^{-3\beta}
\end{equation}
where $r_{c}$ is the core radius and $\beta$ characterizes the power-law
decay (cf. \citet{Beta_model_Sarazin77}). 
Once we have a parametrization
of the emission measure, we solve for the background radius, $r_{bkg}$,
by equating the two area integrals depicted in Figure \ref{BackgroundFig}:
\begin{equation}
Area\, B=4\pi\int_{r_{in}}^{r_{out}}\int_{\sqrt{r_{out}^{2}-r^{2}}}^{R}EM(z,r)\, r\, dzdr
\end{equation}
and area C, 
\begin{equation}
Area\, B\cong Area\, C=2\pi\int_{r_{out}}^{r_{bkg}}\int_{-R}^{R}EM(z,r)\, r\, dzdr
\label{chandra eq deproj}
\end{equation}
where $R$ is the extent of the galaxy emission, the annular area of interest is
from $r_{in}$ to $r_{out}$ with a background from $r_{out}$ to $r_{bkg}$.

With the background methods defined, a set of annuli were extracted
for each galaxy. We choose to bin the data radially into roughly equal
signal-to-noise bins after background subtraction, allowing for a fair comparison
of the statistics in each bin. Because of the demanding number of
parameters in the spectral fitting, we determined five thousand count
bins to provide the best balance between radial sampling and statistical
population of the relevant energies for temperature profiles. For
metallicity determinations, we require ten thousand count bins. In
observations with fewer than five thousand counts, we took a single
annular bin with a size based upon a visual inspection of the extent
of the emission.

Once the regions were defined, the pulse invariant spectra were
extracted from the specified regions with the point sources removed.
The response and effective area matrices were created for each bin
using the extended emission tools mkrmf and mkarf within CIAO. Each
of the source spectra were adaptively regrouped to have a minimum of 25 counts
per channel for improved statistics.

\subsection{\label{Plasma Model}Plasma Model and Spectral Fitting Technique}

For the spectral modeling, which historically produces inconsistent
results when not considered carefully, we build upon our previous
work with \emph{Chandra}, taking extensive measures defining the models
and fitting routines, and perform multiple cross-checks and calibrations to
confirm our methodology. The spectral models were defined and analyses
performed with the goal of obtaining comparative metallicity and temperature
information between the individual galaxies.

Since the hard, unresolved LMXB component acts as an underlying background
to the high energy lines of sulfur, silicon, and magnesium, an accurate
determination of the power-law spectral slope is critical for reliable
metallicity measurements. In a subset of fifteen galaxies of this
sample, we have shown that there appears to be a universal nature to the spectral
properties of early-type galaxy LMXBs that spans three orders of magnitude
in X-ray luminosity \citep*{Jimmy_LMXB}. The errors of the determined
power-law slope of $\Gamma=1.56\pm0.02$ represents an order of magnitude
reduction in the uncertainties from previous studies and allows the
underlying, unresolved LMXB component in our composite spectral modeling to be
fit with a high degree of confidence.

It has been discovered that the lower energies, below 1.0
keV, are undergoing a continually increasing absorption, generally
agreed to be caused from out-gassing of unknown materials in the \emph{Chandra} spacecraft
and subsequent condensation on the blocking filters (e.g. \citet{Chandra_Absorption}).
Of particular interest to this study is the oxygen abundance, obtained
through the {[}O VII{]} and {[}O VIII{]} lines at 0.57 keV and 0.7
keV. Although there is a standard correction for this absorption included
in CIAO, the effectiveness of this correction has yet to be fully
evaluated.

In order to determine the effectiveness of the absorption correction,
we examined the results of this model when performed on the point
sources in galaxies that contained more than 75 detected sources in
an observation. Using the sources has the advantage that the data
are contained within the same dataset as the diffuse emission and
have a much simpler spectral signature of a power-law, with excellent
constraints placed on the slope, as discussed above. In addition to
providing information on the corrected quantum efficiency degradation,
this also allowed us to investigate the well calibrated range of \emph{Chandra}
and the agreement with the literature reported galactic column density
\citep{nH_Dickey_Lockman}. As previously mentioned, we removed likely
background AGN, sources with $L_{X}>2\cdot10^{39}erg\, s^{-1}$, and the central
source, which can be an AGN with different spectral characteristics.
From 0.35 keV to 1.0 keV, residuals less than the Poisson noise
were observed (typically with errors less than 15\% about
the model fit). From 0.284 keV to the carbon edge at 0.35 keV, the
residuals are noticeably large and systematic. 

The Galactic absorption is somewhat degenerate with the blocking filter
absorption as both act to suppress low energy counts. In addition,
the Galactic absorption has influence on the oxygen abundance determination as
this line is well-blended with the iron-L complex and an increase
in the Galactic column density can effectively swamp any signal from
the oxygen lines while leaving the iron abundance unchanged.
In order to minimize the susceptibility of this parameter
on oxygen abundances (note that this parameter has very little impact
on the abundances of the other metals), we decided to fix the 
Galactic absorption in all radial bins. By employing this constraint,
the plasma modeling of the hot gas will preserve relative differences,
while leaving open the possibility of an absolute correction. We determined
this correction to be at most a 30\% correction in oxygen abundance
from a 50\% uncertainty in galactic hydrogen density. For two galaxies,
NGC 720 and NGC 1316, there were large systematic residuals in the
low energy when the standard Galactic column density was applied and
we refit the Galactic column density to values $1.55\cdot10^{20}cm^{-2}$
and $1.18\cdot10^{20}cm^{-2}$, respectively, with successful results.

Based on the point source observations, in addition to tests with
the diffuse gas, we determine the energies from 0.35 keV to 8.0 keV
to be well calibrated. This energy range is employed in all subsequent
spectral analyses.
By extending the spectral
fitting down to 0.35 keV, we find that the temperature-iron metallicity
degeneracy is somewhat relieved and the plasma models can discriminate
between subtle changes in iron abundances for a fixed temperature.

There is substantial evidence indicating that solar metallicity
ratios are not physically representative in early-type galaxies. In
addition to statistically unacceptable fits with common scaling of
solar ratios for all metals, the \emph{XMM-Newton} RGS spectra of
the central region of NGC 4636 and NGC 5044 indisputably reveal different
metallicities for oxygen, iron, magnesium, neon and nitrogen relative
to solar \citep{XMMRGS_NGC4636,XMMRGS_NGC5044}. Also important to
the study of elliptical galaxies, there is no
evidence for multiple temperatures within the central few arcminutes in these RGS spectra.

Motivated by these high quality spectral observations, the
diffuse gas was modeled as plasma represented by the APEC model \citep{APEC}
in which each of the individual elements is allowed to vary (\emph{{}``vapec''}
in XSpec). (We note that the MEKAL model produced similar results.)
Of particular concern was fixing as many of the quantities in the
models as can be scientifically justified, to constrain the solution,
where abundances can behave unpredictably and non-physically when too many
parameters are allowed to vary. 
The bright galaxies were examined and each of the metals
in the plasma code were probed to determine which had negligible impact
upon the solution. If changing a metal content from 0.1 to five times
solar changed the merit of the fit insignificantly, then the metal's
abundance was fixed to the iron abundance, the best determined abundance
in low temperature plasmas. Some of the metals, such as nickel, neon,
and sulfur, had an impact upon the fitting, but did not return realistic
and consistent physical values as judged by the consistency seen in the radial variations of these
quantities. The sulfur line is weak
and there is insufficient spectral resolution to isolate
the nickel and neon lines from the iron-L complex. These metals are
allowed to vary in the modeling, but we do not report the results.

It is not possible to fit the faint galaxies with a variable abundance
model. For galaxies with less than $\sim5000$ gas counts, we employ
the observed consistent metal ratios in the bright galaxies (see Section
\ref{Chandra Results Spectral}) to extend the metallicity determinations
to the faint galaxies. The average metal ratios relative to solar
are locked from the median of the bright galaxies creating a single
parameter abundance for fitting.

To allow for gain shifts, we permitted the redshift to vary for all
the radial bins in a preliminary fit, and a common, average redshift was
frozen for all subsequent modeling.  This offset in the recessional velocity
significantly improves the residuals around the relatively well isolated
Si and Mg lines.  The reason for selecting a common gain shift for all 
the radial bin, instead of allowing this parameter to be fit in each bin,
is that the relative consistency of the fitting from bin to bin is improved with
a common value.

\section{\label{Chandra Results}Spectral Results}

A division is made between the data presentation (this Section) and
a discussion of their implications (forthcoming publications), due
to the extensive nature of spatial and spectral fitting of 54 galaxies
in multiple radial bins. All subsequent errors reported are one-sigma
(68\%) unless otherwise stated. A Hubble Constant of $71\, km\, s^{-1}\, Mpc^{-1}$
is used throughout \citep{HST_Key_Ho,Map_preprint}. Note that this
choice of Hubble constant is consistent with the employed distance
determinations from surface brightness fluctuations (Table \ref{Chandra Sample Table};
\citet{Tonry_SBF_distances}). In the reporting of the metallicities,
we conform to the X-ray standard of using the solar abundances taken
from the photospheric measurements reported in \citet{Abundances_Anders_Grevesse}.
These numbers are immediately comparable to output from the defaults
set in X-ray spectral fitting programs, XSpec and Sherpa.

\subsection{\label{Chandra Results Spectral}Radial Spectral Properties}

The results from the spectral fitting of the radial bins are presented
in Tables \ref{Chandra Radial Table: Temp} - \ref{Chandra Radial Table: Luminosities Outer}.
Many of the faint galaxies only have one radial bin. Table \ref{Chandra Radial Table: Temp}
is based on the
nominally 5000 counts per bin regions while the galaxy mass and luminosity information, 
in Tables \ref{Chandra Radial Table: Mass}, 
and \ref{Chandra Radial Table: Luminosities Outer},
as well as metallicity information
presented in Table \ref{Chandra Radial Table: Z} and \ref{Chandra Radial Table: faint Z},
are based on 10,000 counts per bin regions.

Table \ref{Chandra Radial Table: Temp} presents radial temperature
determinations and 1$\sigma$ errors for each of the three backgrounds
defined in Section \ref{Chandra Backgrounds and Regions}. Galaxies
with three or more radial bins are also presented graphically in Figures
\ref{ktzchi_NGC507} - \ref{ktzchi_NGC5846}.

Table \ref{Chandra Radial Table: Mass} presents radial density, entropy,
cumulative mass and the fraction of area sampled by the S3 chip for
the prescribed annular bin. The data in this table are from spectral
fitting with an outer sky background. Using a blank background region
produced similar results. Galaxies
with three or more radial bins are also presented graphically in Figures
\ref{rhomassE_NGC507} - \ref{rhomassE_NGC5846}.

Table \ref{Chandra Radial Table: Z} presents individual elemental
abundances for oxygen, iron, silicon and magnesium in radial bins
for bright galaxies. Galaxies with three or more temperature radial bins are also
presented graphically in the middle panel of Figures \ref{ktzchi_NGC507}
- \ref{ktzchi_NGC5846}.

\begin{figure}
\epsscale{0.75}
\plottwo{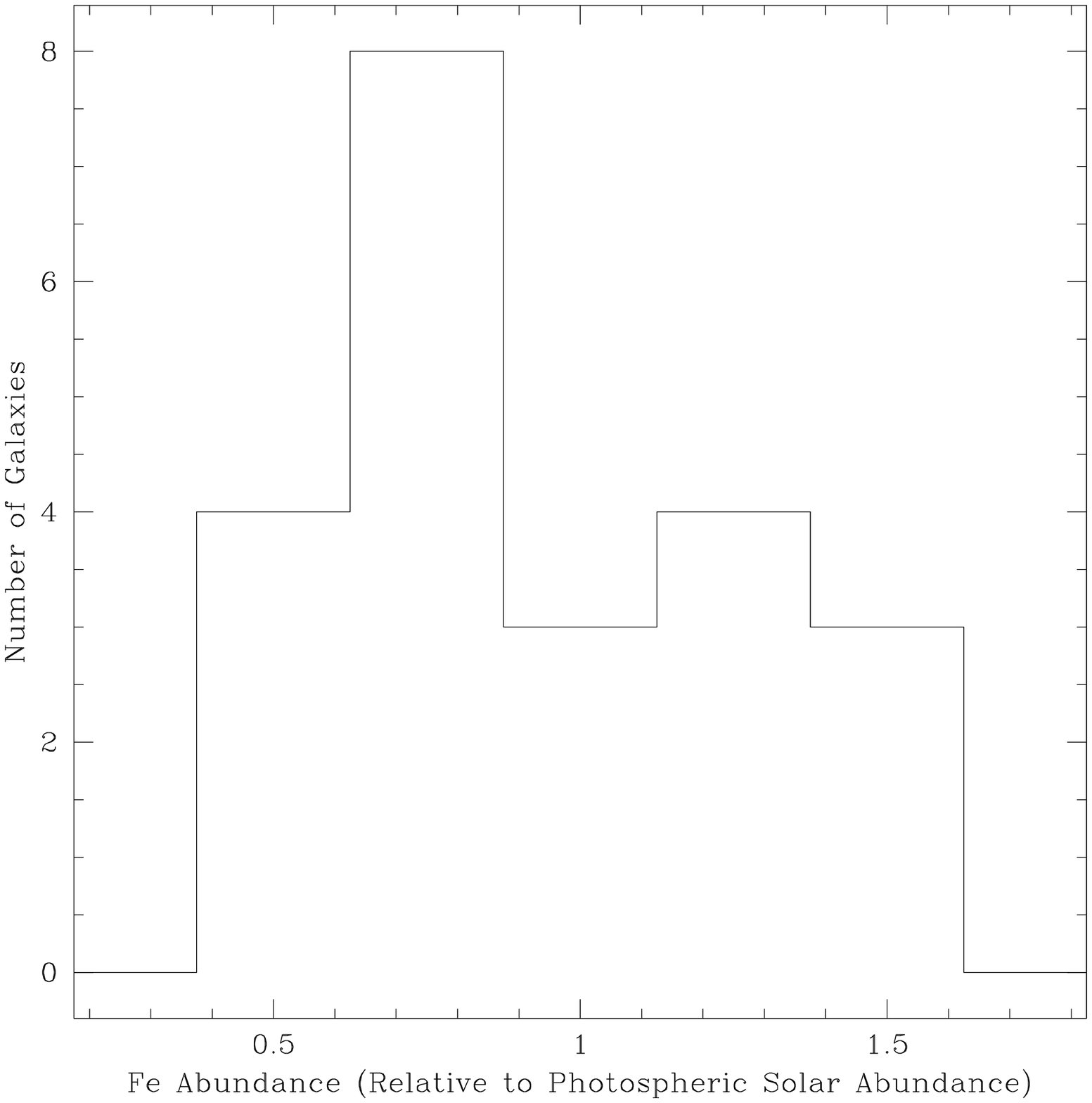}{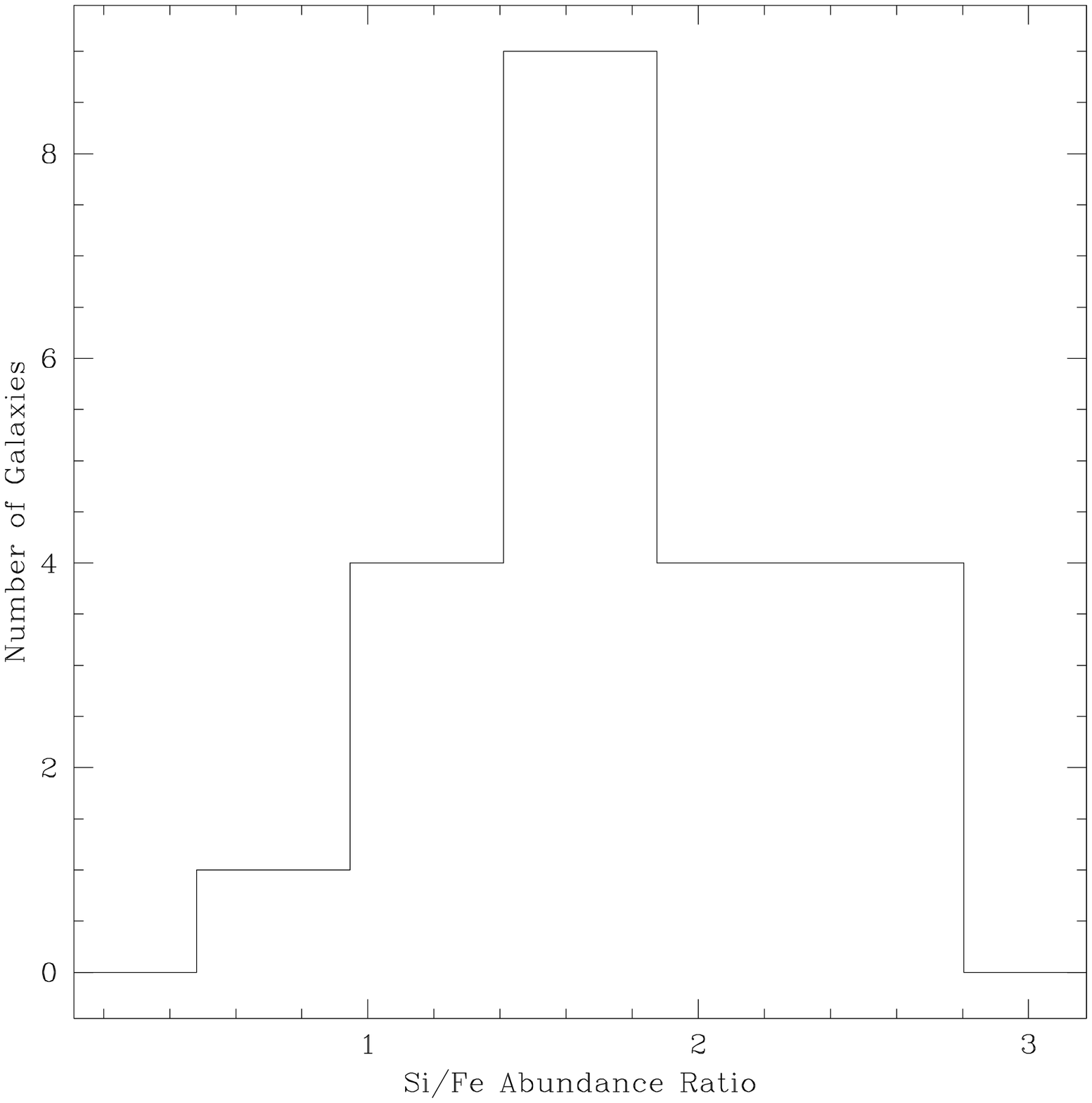}
\caption[Histogram of Metals Determined from the X-ray Gas: Iron and Silicon]
{Histogram of bright X-ray galaxies where signal-to-noise was sufficient for individual
elemental abundance determinations.  On left is the distribution of iron metallicity for the
bright galaxies and on the right is the silicon to iron ratio.}
\label{Chandra Bright Z Figs 1}
\end{figure}

\begin{figure}
\plottwo{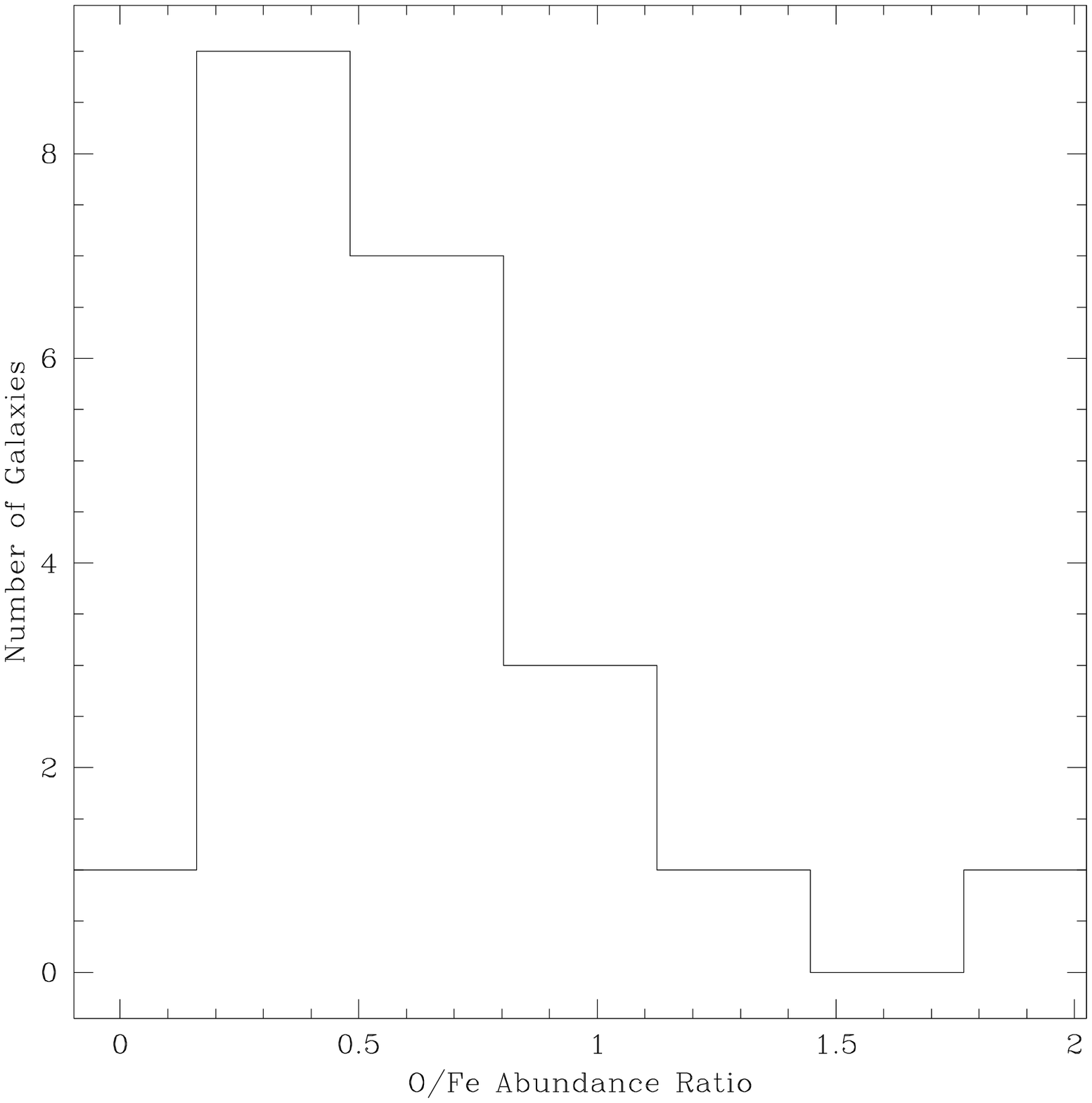}{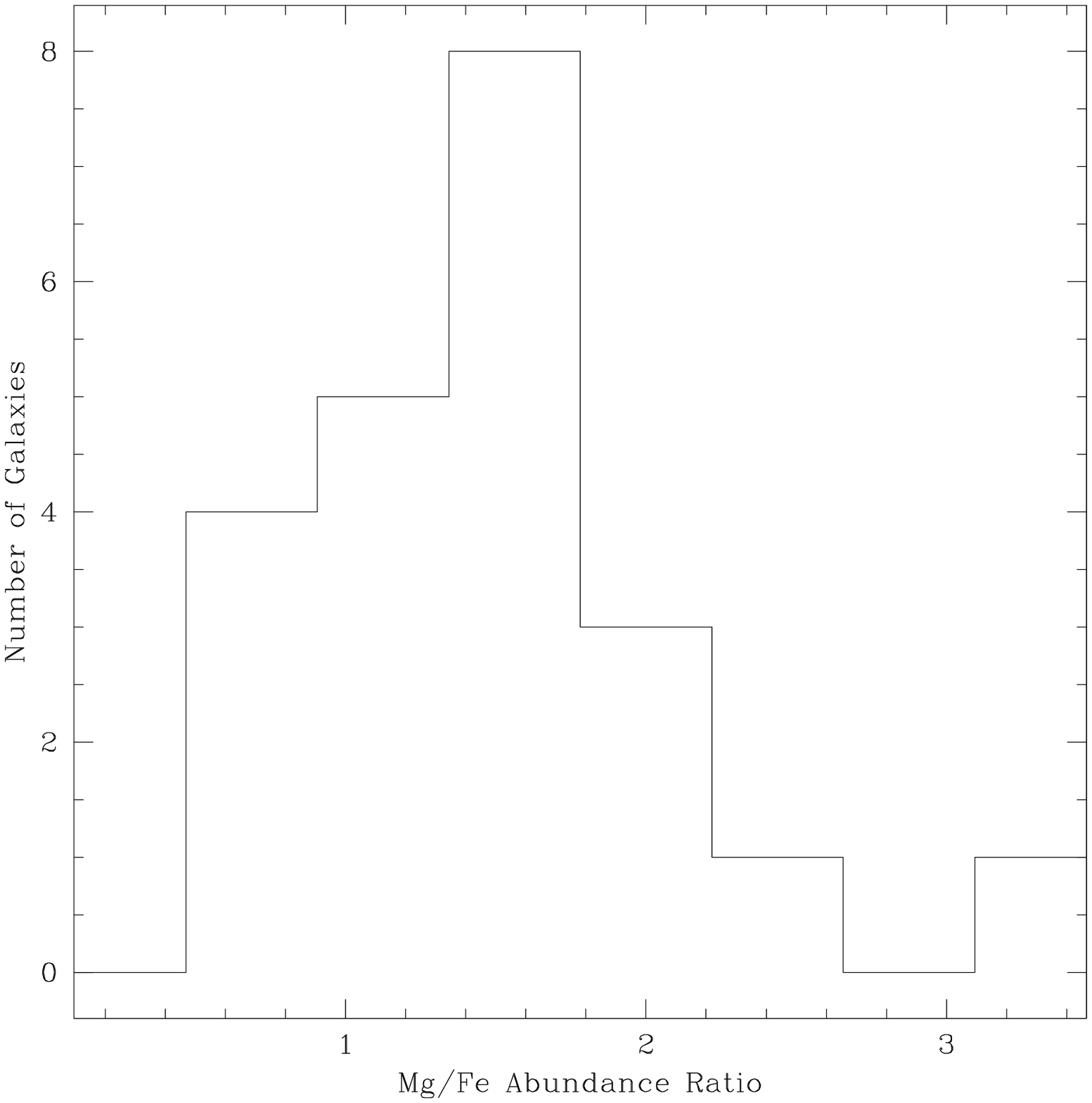}
\caption[Histogram of Metals Determined from the X-ray Gas: Oxygen and Magnesium]
{Histogram of bright X-ray galaxies where signal-to-noise was sufficient for individual
elemental abundance determinations.  On left is the distribution of oxygen over iron 
metallicity for the bright galaxies and on the right is magnesium to iron ratio.}
\label{Chandra Bright Z Figs 2}
\end{figure}

Histograms of the metal content in the high signal-to-noise galaxy
observations are shown in Figures \ref{Chandra Bright Z Figs 1} and
\ref{Chandra Bright Z Figs 2}. Although there is a fairly large range
of iron abundances observed, the ratios of iron to the other independently
determined elements of silicon, oxygen and magnesium are narrowly
distributed. This observation indicates that early-type galaxies have
common enrichment histories, and, further, the non-unity values of
these ratios reveals that this ISM enrichment process is different
than that which contributed to production of the solar metal
content. These ratios can be understood as a gas that has had
its enrichment dominated by SN Ia's. 

Employing these results, we fit the low signal-to-noise early-type
galaxy observations with a single abundance, but requiring abundance
ratios of $Si/Fe=1.65$, $O/Fe=0.35$, $Mg/Fe=1.55$, and all other
elements equal in abundance to iron. The results from the metallicity
fitting of the low signal-to-noise subsample are presented in Table
\ref{Chandra Radial Table: faint Z} as an iron abundance when
the metals are tied together as prescribed above.

The unresolved binary component, $L_{X}^{usrc}$ compensates for different
background levels and thus significant differences are observed in
this component between the blank sky subtractions and outer region
background. The gas and resolved source luminosities are largely unaffected.
It is not entirely clear which background is more representative for
a given galaxy and will depend on physical conditions of the local
galaxy environment and location in the sky. It is likely that in NGC
1399, when the blank sky subtraction is used the $L_{X}^{usrc}$ component
is tracing not only the unresolved LMXBs but also the underlying hot
$kT\sim3keV$ ICM can be seen in the outer bins. However, subtracting
an outer region from a galaxy that fills the S3 chip, such as NGC
4472, over estimates the background since the background region is
contaminated with galaxy emission. Further complicating matters, \citet{Chandra_Absorption},
in study of the sky positional dependence of the soft X-ray background, find significantly
different emission levels for the background in addition to differences
in the spectra shapes of the background. These observations suggest
that the background should be taken locally if possible.

The background choice for bright galaxies defines the selection
of the physical components analyzed in the spectral analysis; With
a blank sky background the ICM, ISM and unresolved LMXB components
are selected while an outer region background selects the ISM and
and unresolved LMXB with perhaps a slight over estimation of the background.
Since the galactic emission is our primary interest, we use an outer
region background for the global property studies (Section \ref{Chandra Results Cumulative}).
We report the luminosity determinations for 
an outer region subtraction, but caution on the accuracy of the
unresolved component for bright galaxies. For faint galaxies, the
outer region background is heavily preferred to the blank sky subtracted
results since the background region is observed through the same Galactic
column density.

Cumulative gas luminosities as well as luminosities per area are presented
in Table \ref{Chandra Radial Table: Luminosities Outer}. Also presented
are the unresolved LMXB and resolved LMXB components. In this table
an outer, on-chip region is used as a background.


Galaxies with three or more radial bins (i.e. Table \ref{Chandra Radial Table: Luminosities Outer}) are also presented graphically in Figures \ref{Lxc_NGC507} - \ref{Lxc_NGC5846}.


\subsection{\label{Chandra Results Cumulative}Cumulative Galaxy Properties}

The cumulative X-ray properties of the sample are derived from the
radial properties presented in Section \ref{Chandra Results Spectral}
and listed in Table \ref{Chandra Composite}.  An outer, on-chip background 
is used and all quantities are scaled to 1 Re.

\section{\label{ROSAT Comparison}Comparison with \emph{ROSAT}}

The gas luminosities detected with \emph{Chandra} are compared to
luminosities determined with \emph{ROSAT} in order to check calibrations
and to look for systematic residuals. We expect X-ray
bright galaxies to be systematically fainter with \emph{Chandra}
ACIS-S3 observations as the limited field of view precludes total
luminosities from being measured. For X-ray faint galaxies, the \emph{Chandra}
data will better isolate the LMXB contribution and it is expected
that the \emph{Chandra} gas luminosities will be lower than the \emph{ROSAT}
ones.

\begin{figure}
\epsscale{0.70}
\plotone{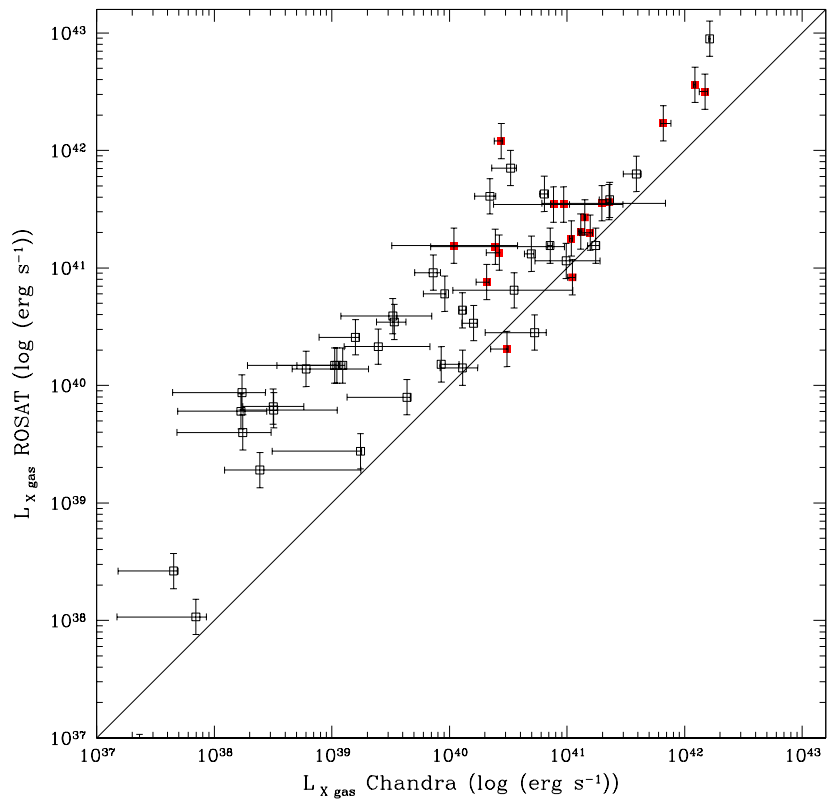}
\caption[Comparison of  \emph{Chandra} and \emph{ROSAT} Gas  Luminosities]
{Comparison between \emph{Chandra} (Table \ref{Chandra Sample Table}) and \emph{ROSAT} gas luminosities.
The \emph{ROSAT} gas luminosities have removed the
stellar binary population from the total X-ray luminosity via statistical 
subtraction based on blue luminosity of the galaxy.
The solid line represents a 1:1 relation.  The filled squares are the 
brightest group galaxies, which have been shown to have their 
gas properties traced best by characteristics of the group rather than the galaxy.
}

\label{Lxcomp}
\epsscale{1.0}
\end{figure}

In Figure \ref{Lxcomp}, \emph{ROSAT} gas luminosities are plotted against the one effective radius gas
luminosities reported in Table \ref{Chandra Composite}.
The \emph{Chandra} luminosities are systematically under-luminous due to 
different scaling radii (typically, \emph{ROSAT} luminosities are scaled to $4\cdot r_{e}$).
The effect of these different scaling radii produce a vertical offset.
Once this offset is applied, the agreement is reasonable as can be seen by the solid
line representing a 1:1 relationship. 
At the X-ray
faint end, the \emph{ROSAT} luminosities are systematically 
brighter than \emph{Chandra} luminosities due to the under-subtraction
of LMXBs. Some of the scatter in this plot is caused by different
distances being employed; although we have corrected for differences
in Hubble constants, the majority of our distances are derived from
surface brightness fluctuations while the majority of \emph{ROSAT}
luminosities were derived distances from radial velocity measurements.
Finally, the \emph{ROSAT} luminosities are not a truly uniform sample
as they have been taken from different authors employing different
processing techniques, spectral modeling and distance determinations.
Given all these factors, the level of scatter in this comparison is
acceptable and, further, we conclude that our data analysis and the
current state of \emph{Chandra} calibrations satisfactorily produce
luminosities consistent with previous X-ray missions.

\begin{figure}
\epsscale{0.70}
\plotone{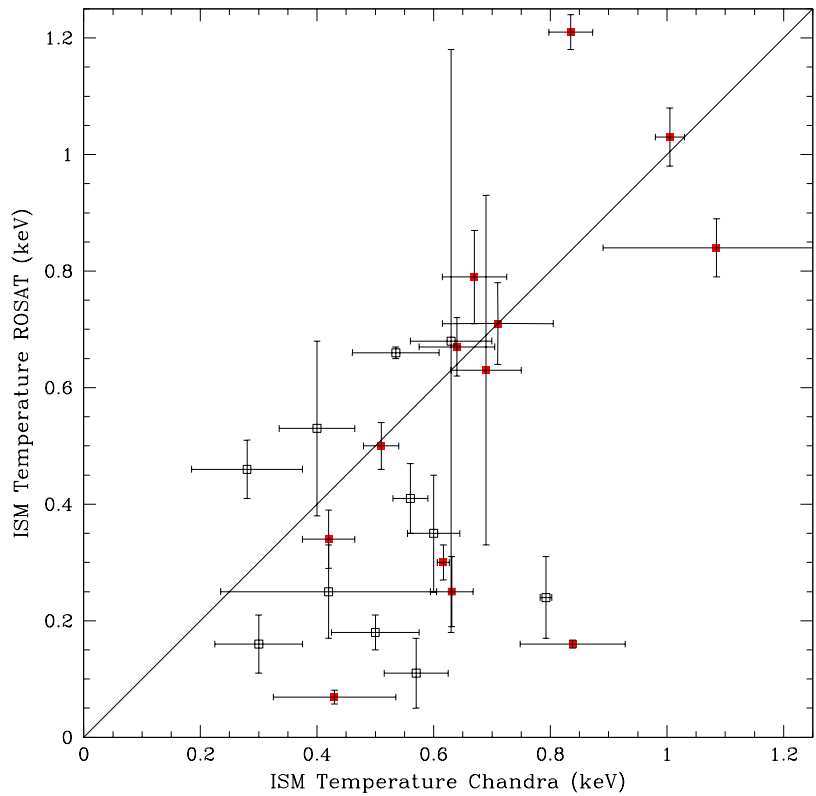}
\caption[Comparison of \emph{Chandra} and \emph{ROSAT} Gas Temperatures]
{Comparison between\emph{Chandra} (Table \ref{Chandra Sample Table}) and \emph{ROSAT} gas temperatures.  The solid line represents a 1:1 relation.  No correlation is observed between the different temperature determinations.   
}
\label{ktcomp}
\epsscale{1.0}
\end{figure}

\begin{figure}
\epsscale{0.80}
\plotone{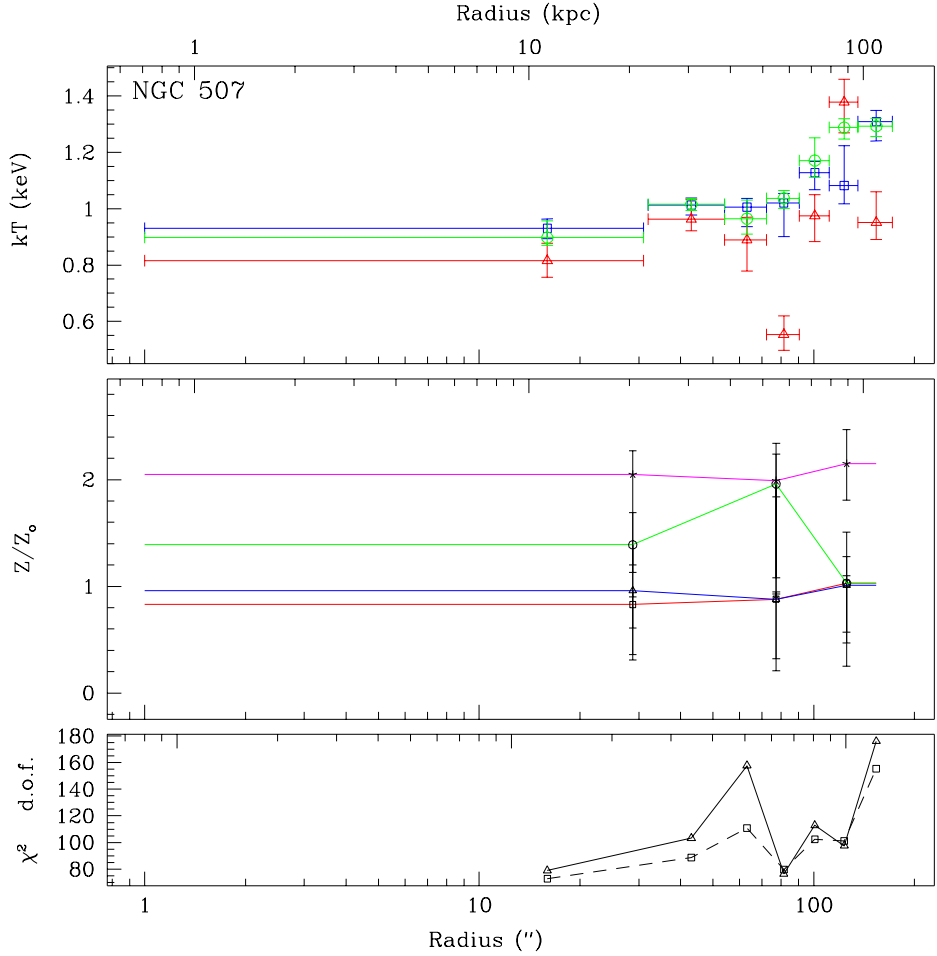}
\caption[Temperature and Metallicity Profile: NGC 507]
{Temperature and Metallicity Profile for NGC 507. 
All panels show radius in arseconds on the lower x-axis and radius in 
kiloparsecs on the upper x-axis.  The top panel displays temperature determinations
with squares representing and on-chip, outer region background, circles 
representing a blank sky background, and triangles representing a deprojection
background (See section \ref{Chandra Backgrounds and Regions} for full 
descriptions of these backgrounds.)  The middle panel displays metallicity 
information with all metals relative to solar photospheric abundances as
reported in \citet{Abundances_Anders_Grevesse}.  In the middle panel
iron abundance is represented as a red square, oxygen abundances is represented 
as a blue triangle, magnesium abundances is represented as a green circle, and 
silicon abundances is represented as a magenta star.  The data in the middle panel
are determined from a on-chip outer region background.
The bottom panel displays the statistical characterization of the model fits
to the data from the on-chip, outer
region background. Typically, the model fits with the blank sky 
background display similar behavior.  The number of degrees of freedom
are represented by squares and the triangles represent chi-squared.
}
\label{ktzchi_NGC507}
\end{figure}

\begin{figure}
\epsscale{0.80}
\plotone{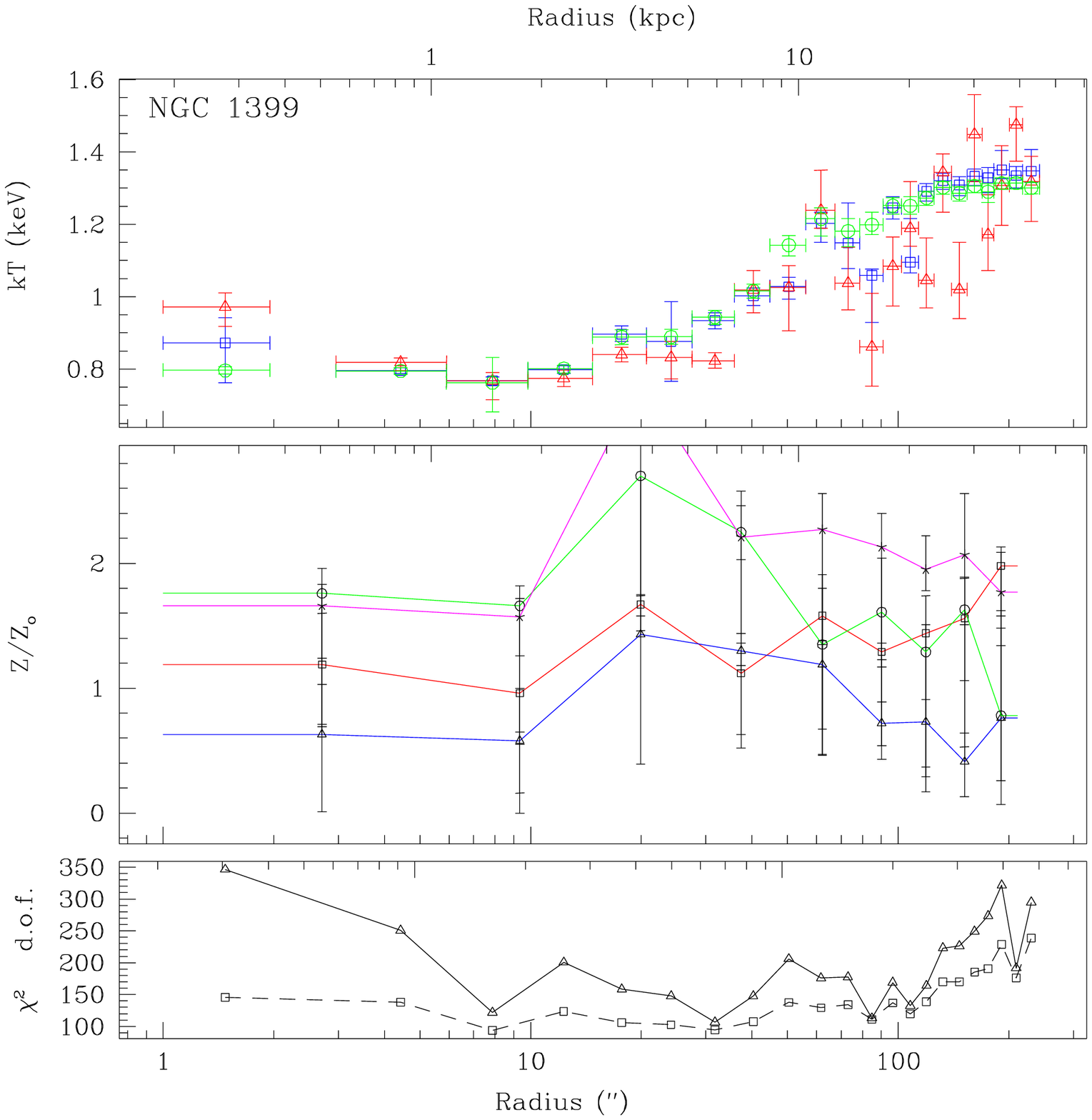}
\caption[Temperature and Metallicity Profile: NGC 1399]
{Temperature and Metallicity Profile for NGC 1399. Same as Figure \ref{ktzchi_NGC507}.}
\label{ktzchi_NGC1399}
\end{figure}

\begin{figure}
\epsscale{0.80}
\plotone{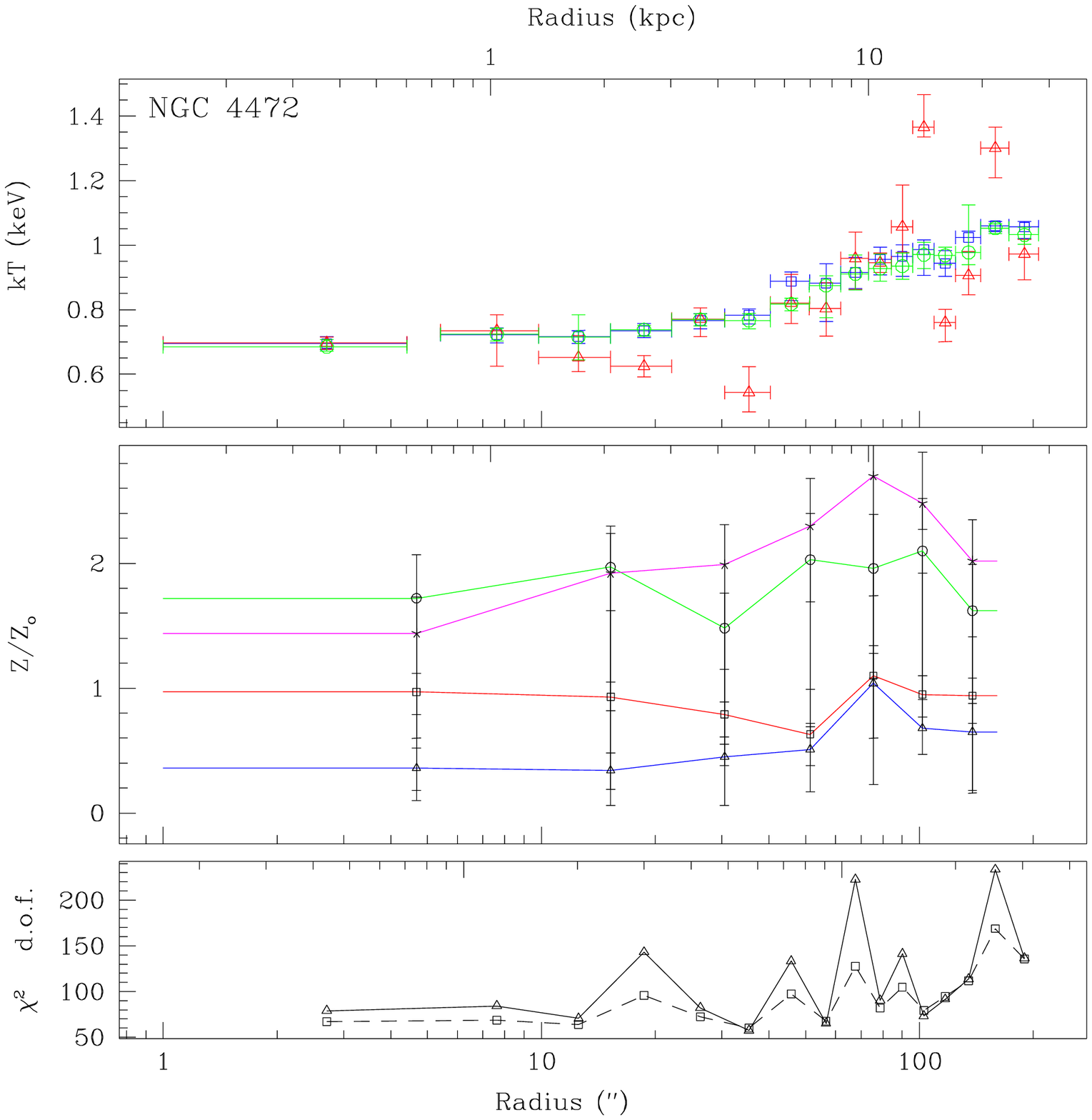}
\caption[Temperature and Metallicity Profile: NGC 4472]
{Temperature and Metallicity Profile for NGC 4472. Same as Figure \ref{ktzchi_NGC507}.}
\label{ktzchi_NGC4472}
\end{figure}

\begin{figure}
\epsscale{0.80}
\plotone{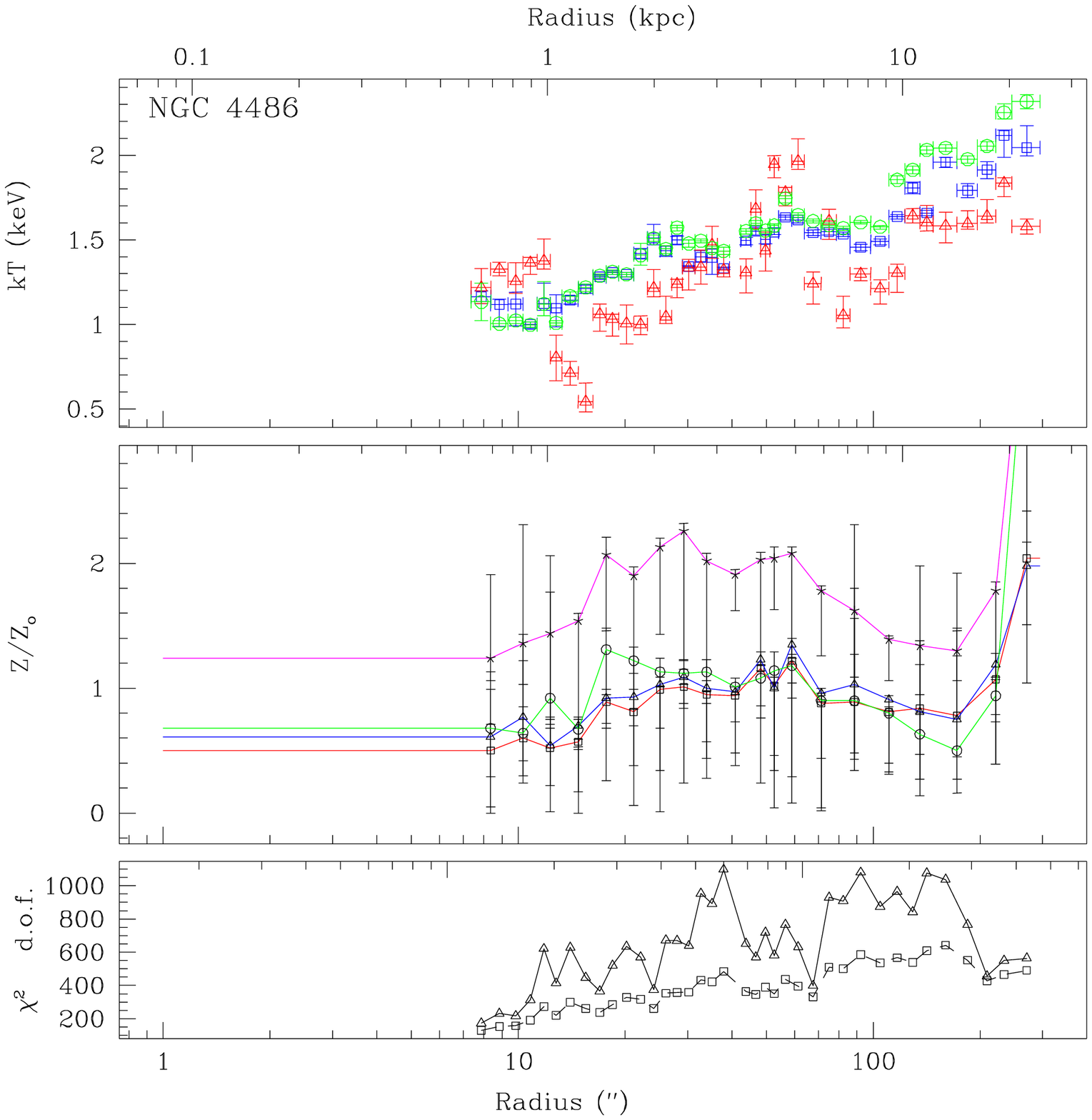}
\caption[Temperature and Metallicity Profile: NGC 4486]
{Temperature and Metallicity Profile for NGC 4486. Same as Figure \ref{ktzchi_NGC507}.}
\label{ktzchi_NGC4486}
\end{figure}

\begin{figure}
\epsscale{0.80}
\plotone{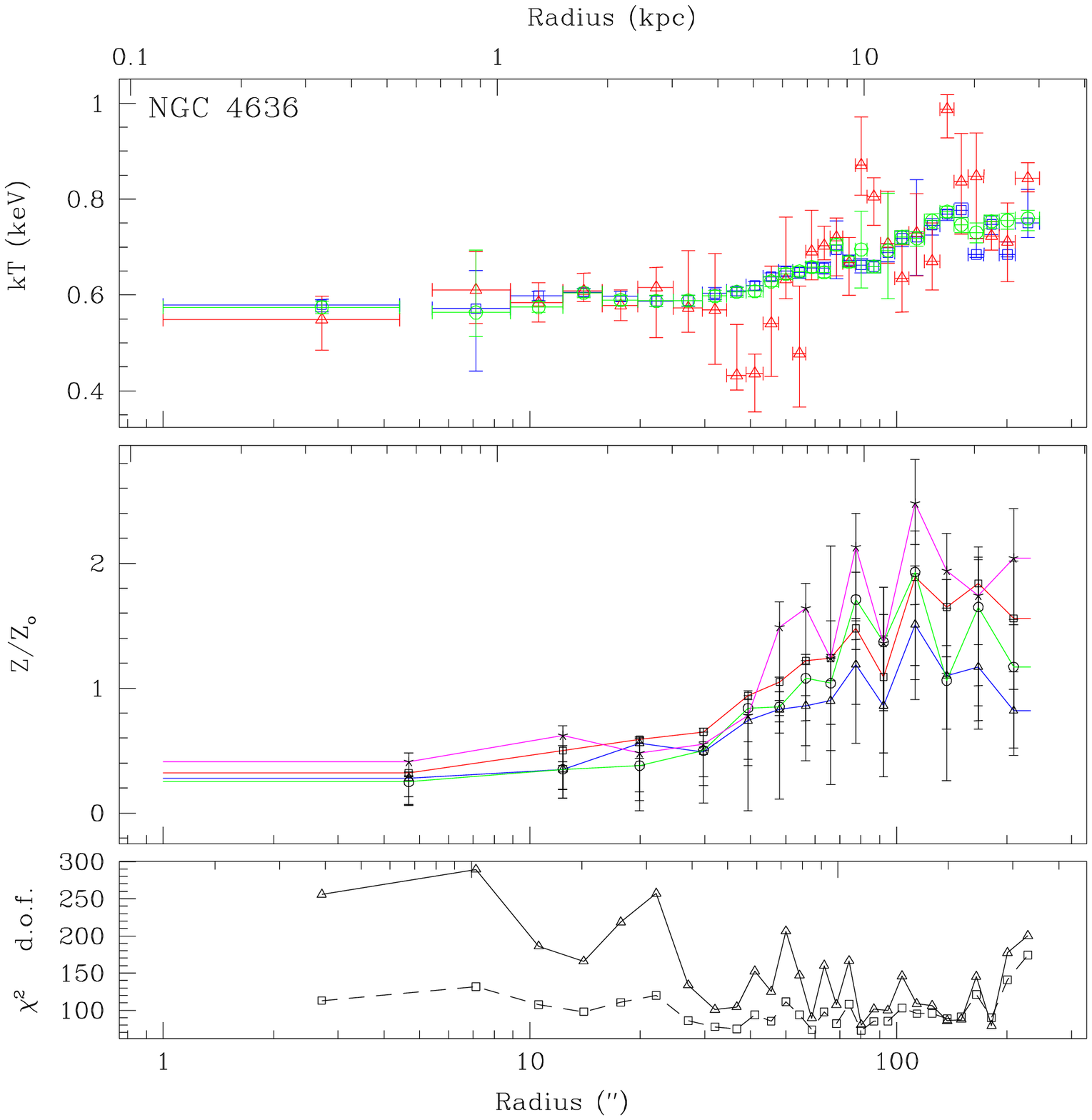}
\caption[Temperature and Metallicity Profile: NGC 4636]
{Temperature and Metallicity Profile for NGC 4636. Same as Figure \ref{ktzchi_NGC507}.}
\label{ktzchi_NGC4636}
\end{figure}

\begin{figure}
\epsscale{0.80}
\plotone{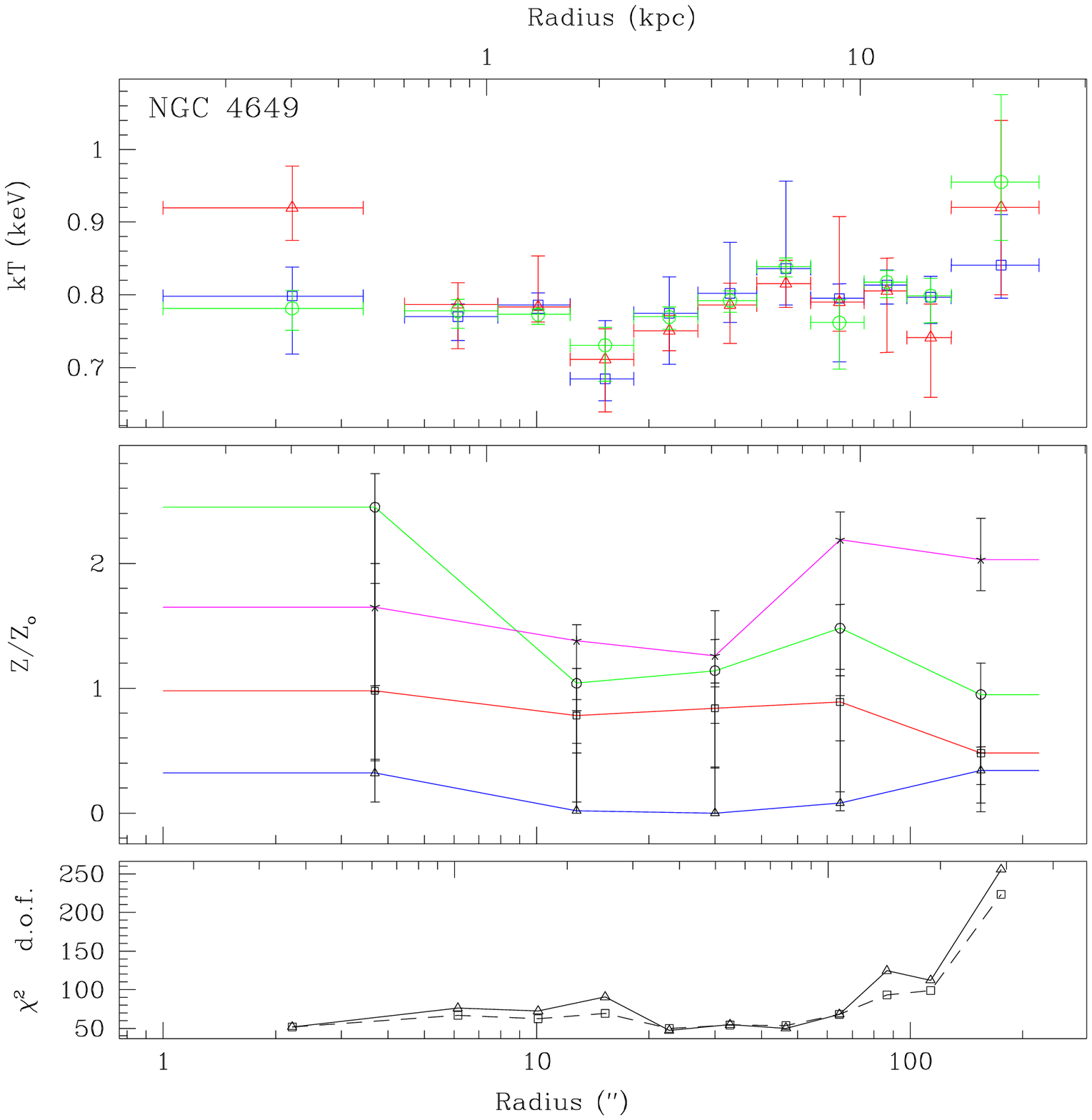}
\caption[Temperature and Metallicity Profile: NGC 4649]
{Temperature and Metallicity Profile for NGC 4649. Same as Figure \ref{ktzchi_NGC507}.}
\label{ktzchi_NGC4649}
\end{figure}

\begin{figure}
\epsscale{0.80}
\plotone{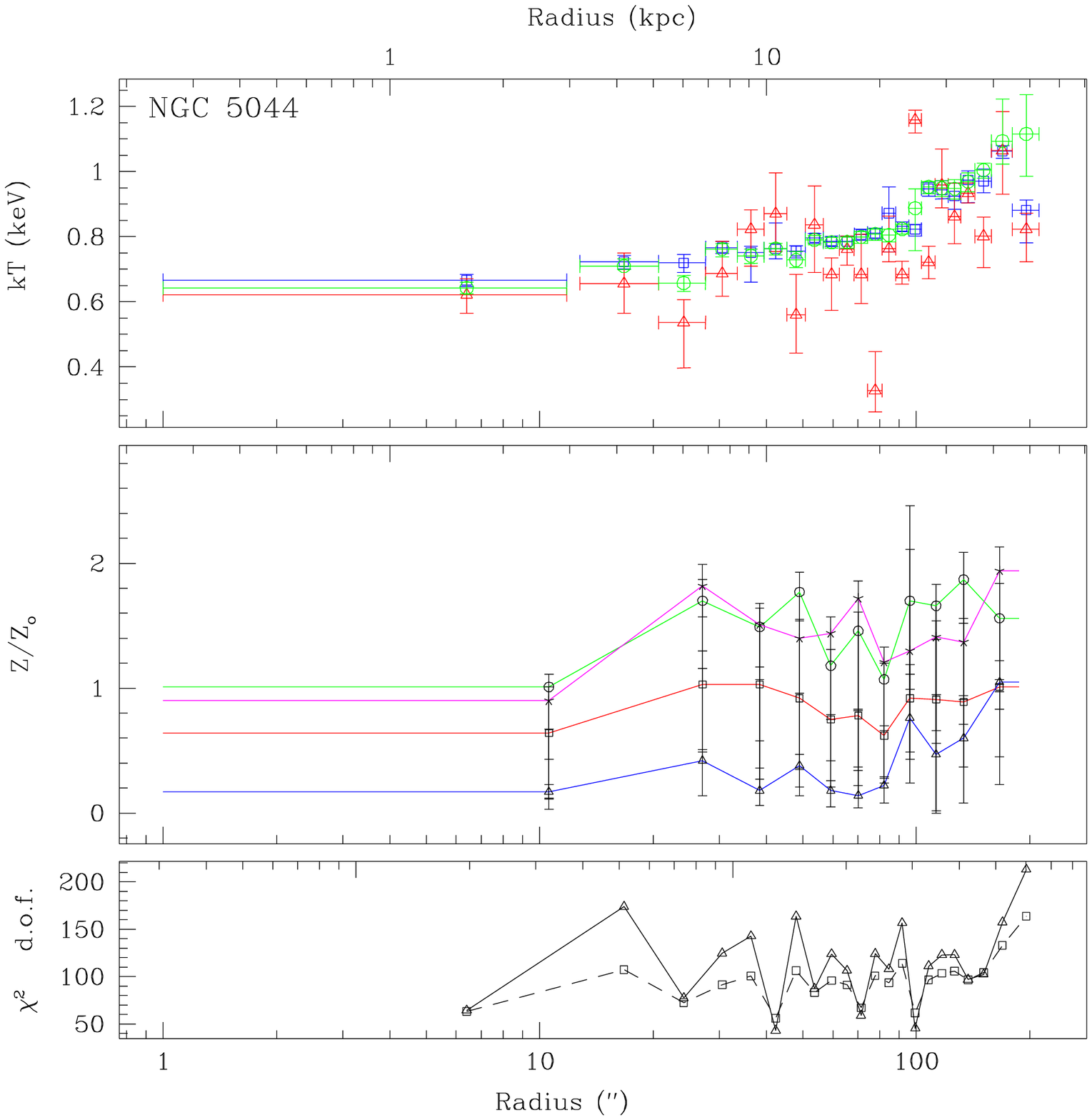}
\caption[Temperature and Metallicity Profile: NGC 5044]
{Temperature and Metallicity Profile for NGC 5044. Same as Figure \ref{ktzchi_NGC507}.}
\label{ktzchi_NGC5044}
\end{figure}

\begin{figure}
\epsscale{0.80}
\plotone{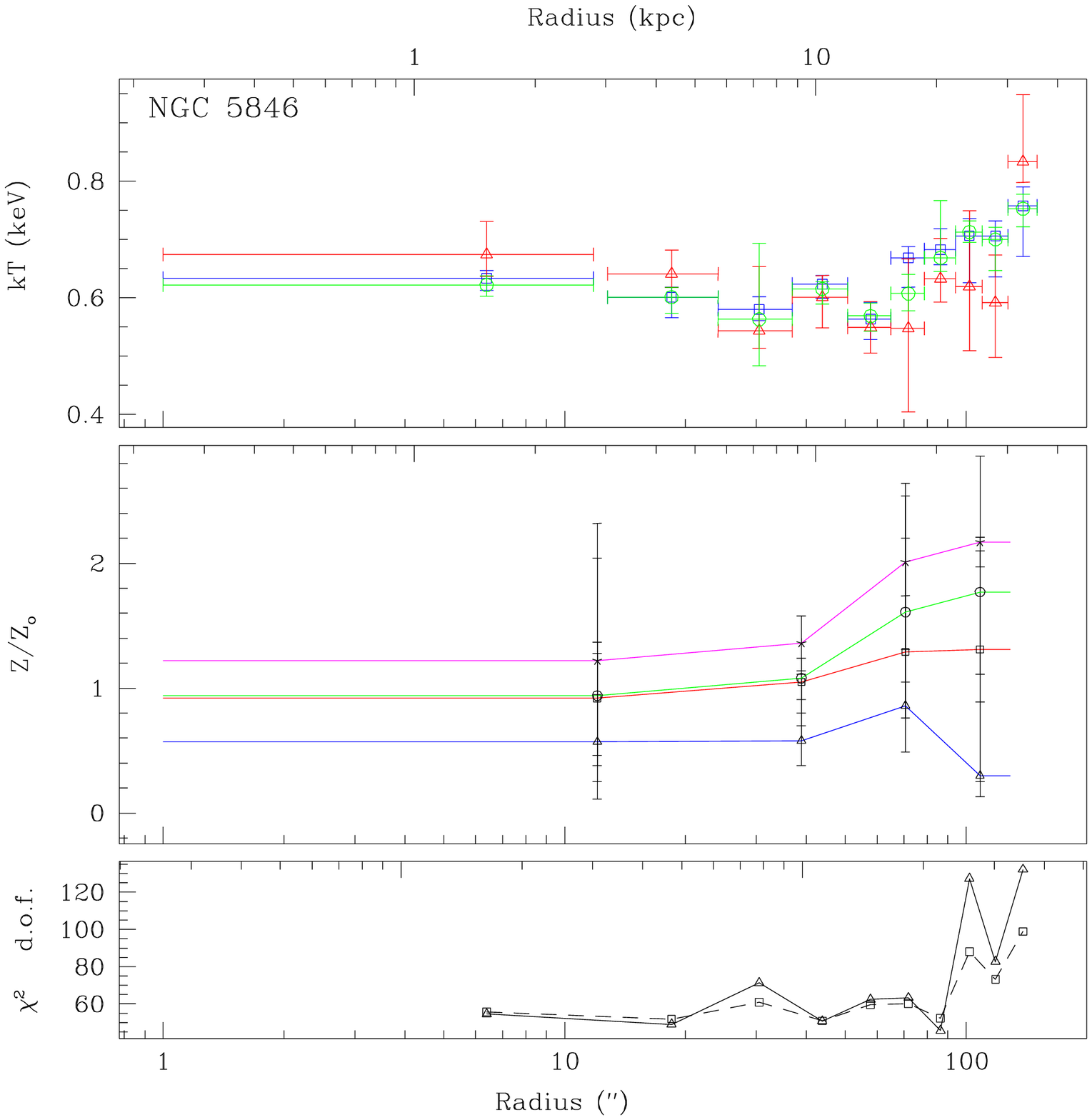}
\caption[Temperature and Metallicity Profile: NGC 5846]
{Temperature and Metallicity Profile for NGC 5846. Same as Figure \ref{ktzchi_NGC507}.}
\label{ktzchi_NGC5846}
\end{figure}

\begin{figure}
\epsscale{0.80}
\plotone{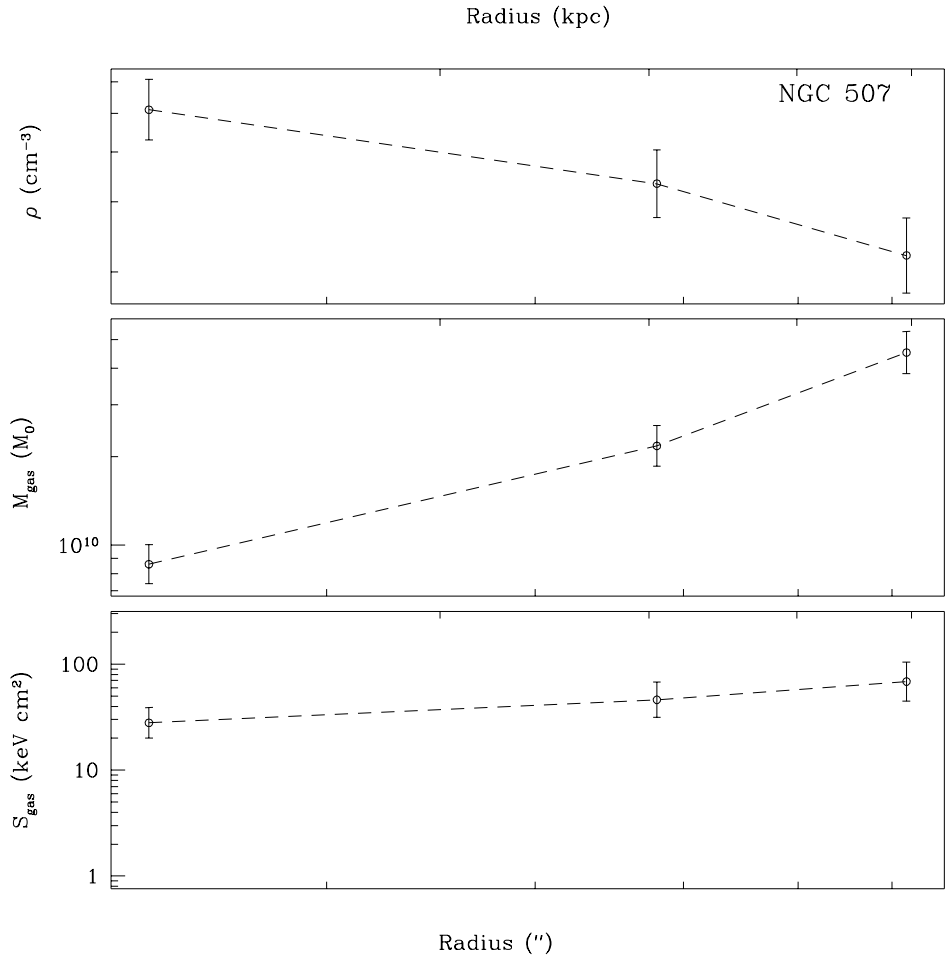}
\caption[Density, Mass, Entropy Profile: NGC 507]
{Density, Mass and Entropy Profile for NGC 507. 
All panels show radius in arseconds on the lower x-axis and radius in 
kiloparsecs on the upper x-axis.  The vertical dotted line shows 
the optical half-light radius.
The top panel displays electron density.  The middle panel displays
the cumulative mass profile in solar units.  The bottom panel displays
the entropy of the gas in units of $ke\, V\, cm^2$.
}
\label{rhomassE_NGC507}
\end{figure}

\begin{figure}
\epsscale{0.80}
\plotone{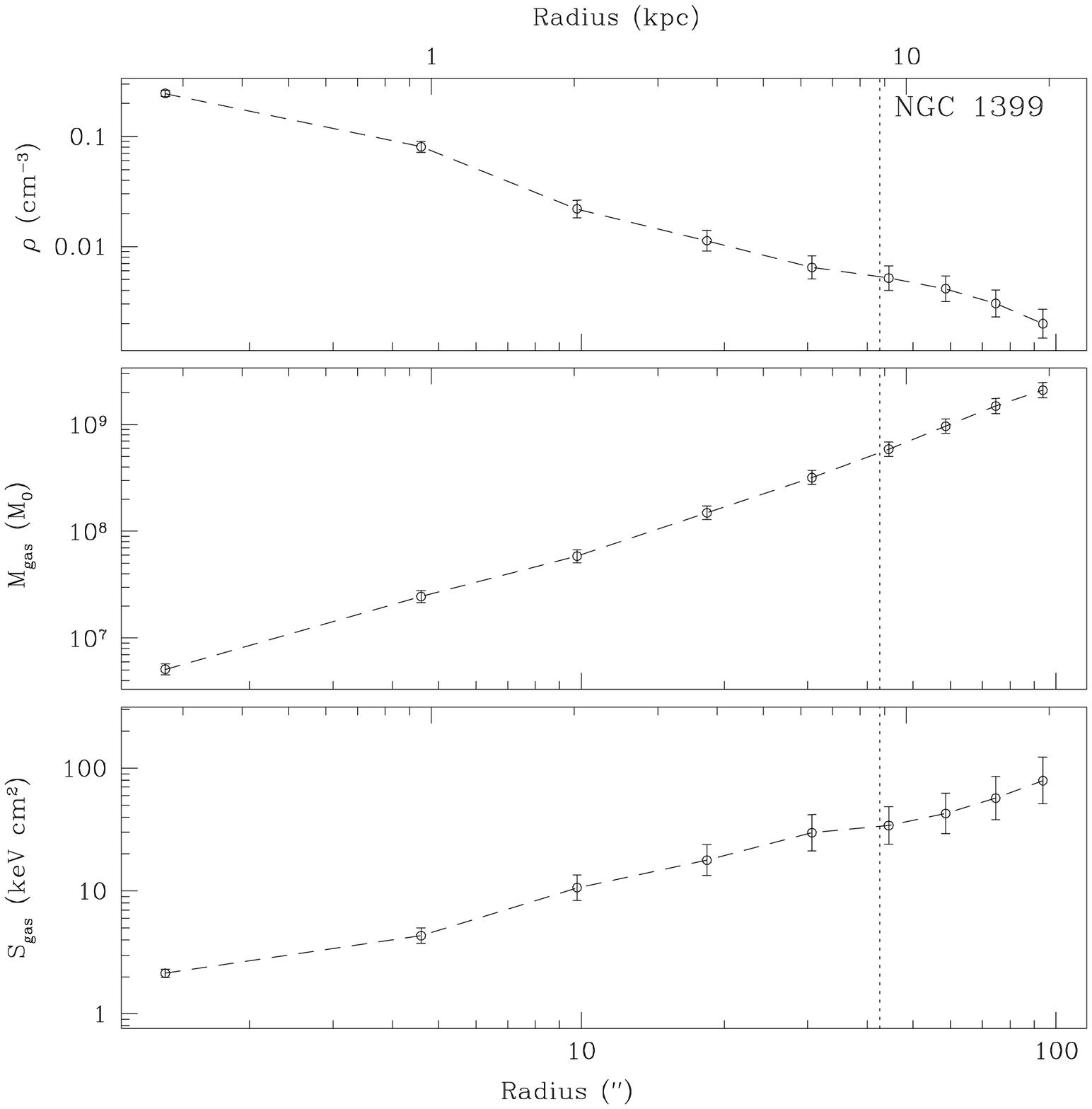}
\caption[Density, Mass, Entropy Profile: NGC 1399]
{Density, Mass and Entropy Profile for NGC 1399. Same as Figure \ref{rhomassE_NGC507}.}
\label{rhomassE_NGC1399}
\end{figure}

\begin{figure}
\epsscale{0.80}
\plotone{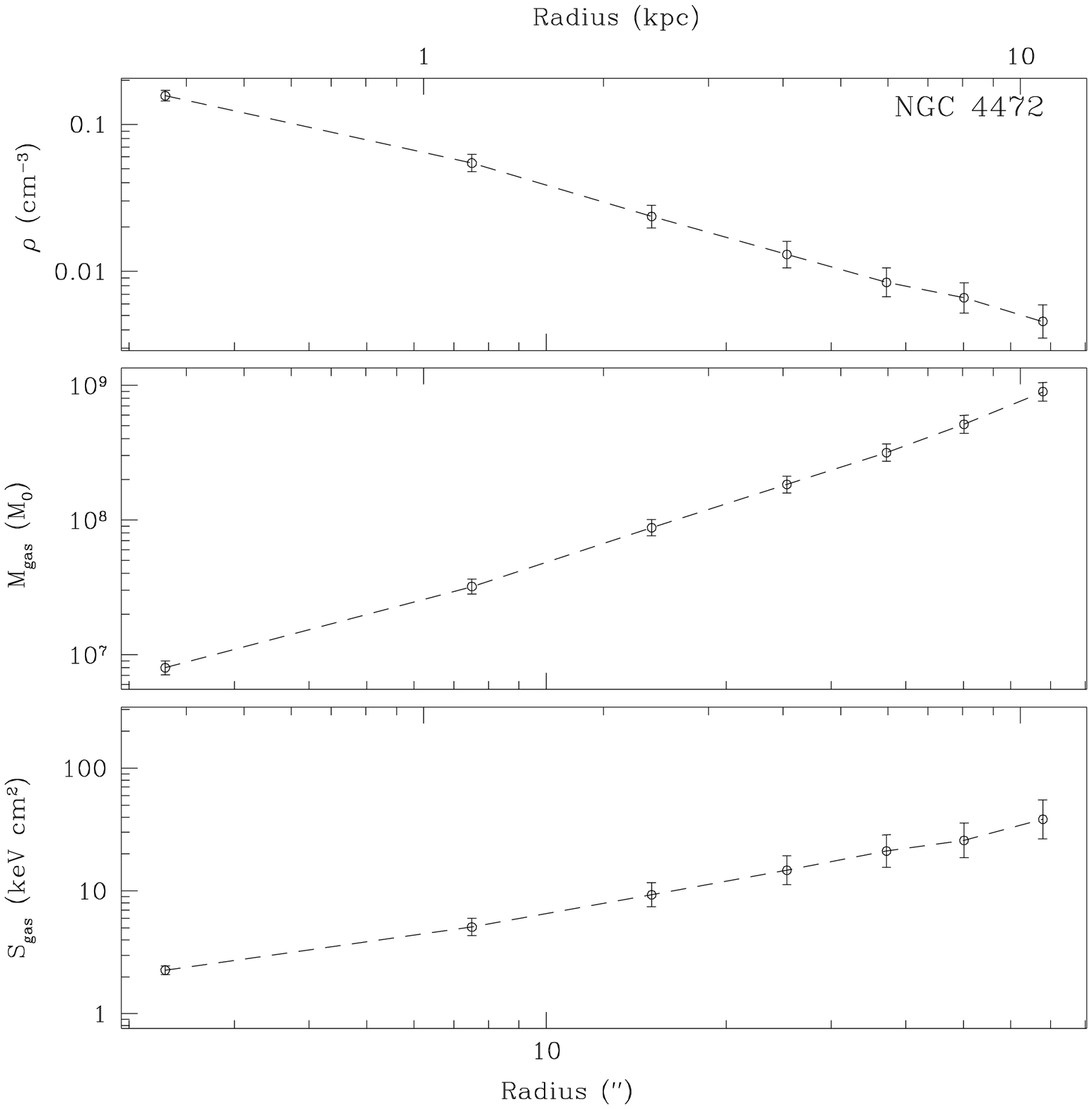}
\caption[Density, Mass, Entropy Profile: NGC 4472]
{Density, Mass and Entropy Profile for NGC 4472. Same as Figure \ref{rhomassE_NGC507}.}
\label{rhomassE_NGC4472}
\end{figure}

\begin{figure}
\epsscale{0.80}
\plotone{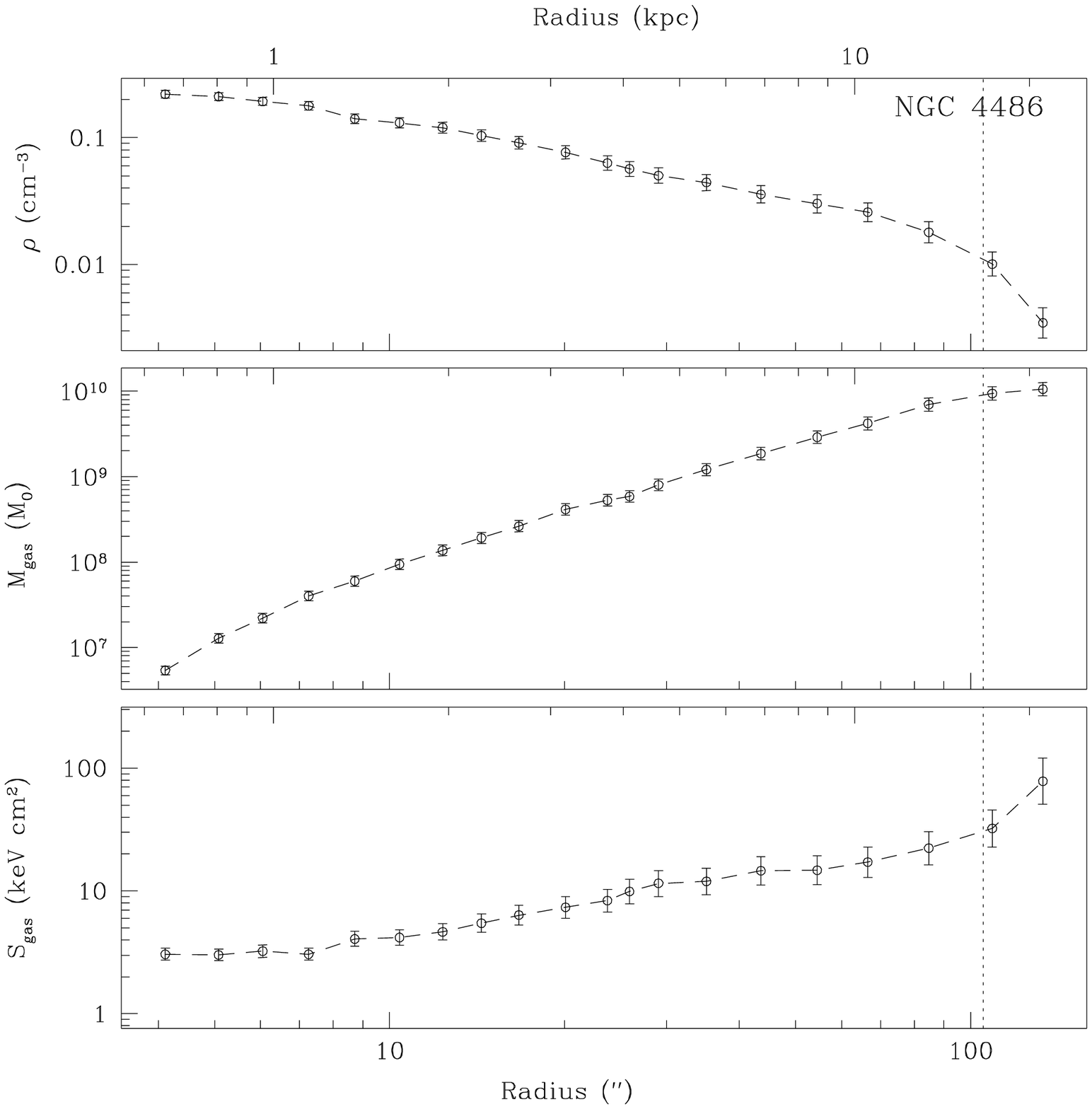}
\caption[Density, Mass, Entropy Profile: NGC 4486]
{Density, Mass and Entropy Profile for NGC 4486. Same as Figure \ref{rhomassE_NGC507}.}
\label{rhomassE_NGC4486}
\end{figure}

\begin{figure}
\epsscale{0.80}
\plotone{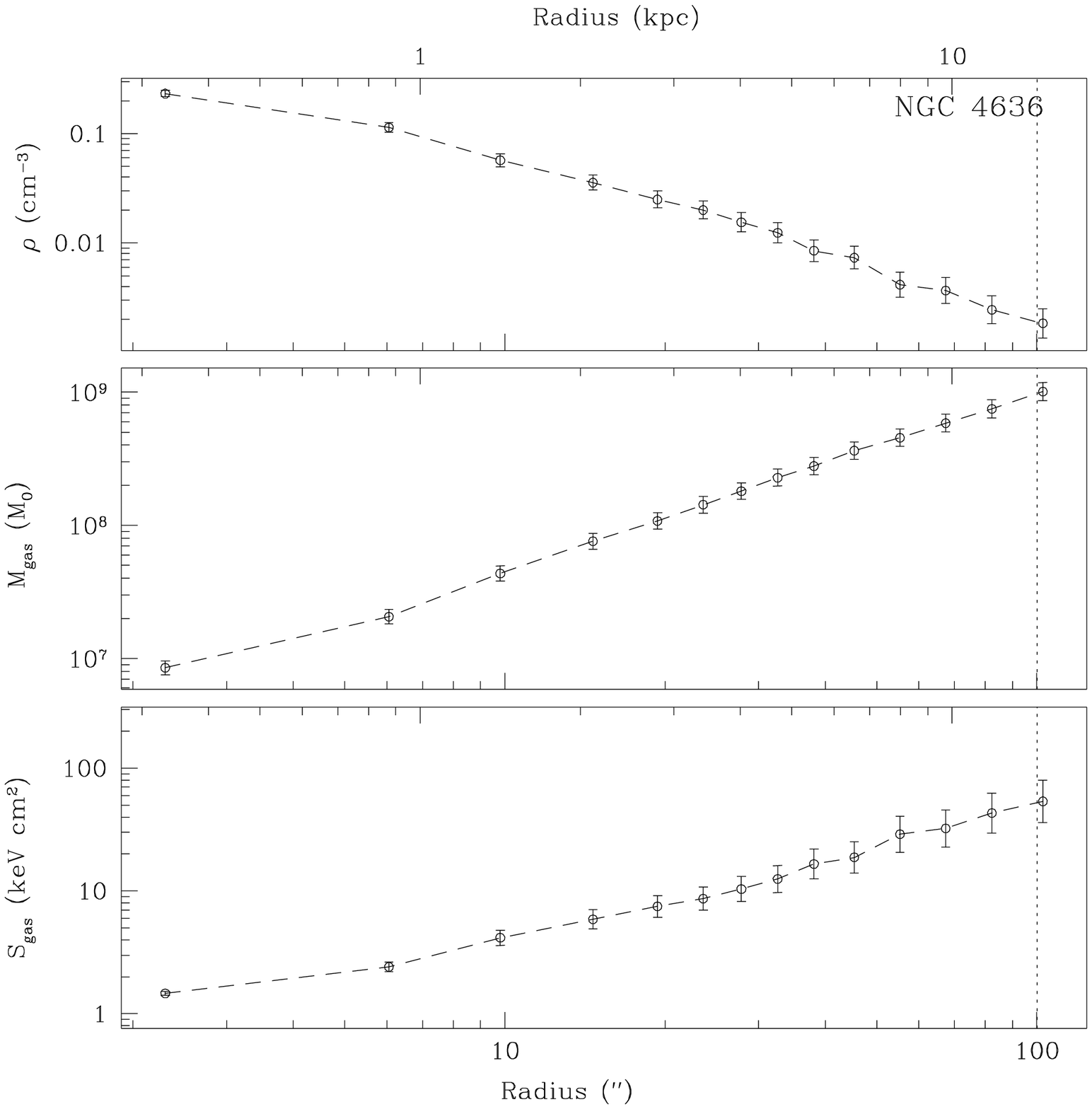}
\caption[Density, Mass, Entropy Profile: NGC 4636]
{Density, Mass and Entropy Profile for NGC 4636. Same as Figure \ref{rhomassE_NGC507}.}
\label{rhomassE_NGC4636}
\end{figure}

\begin{figure}
\epsscale{0.80}
\plotone{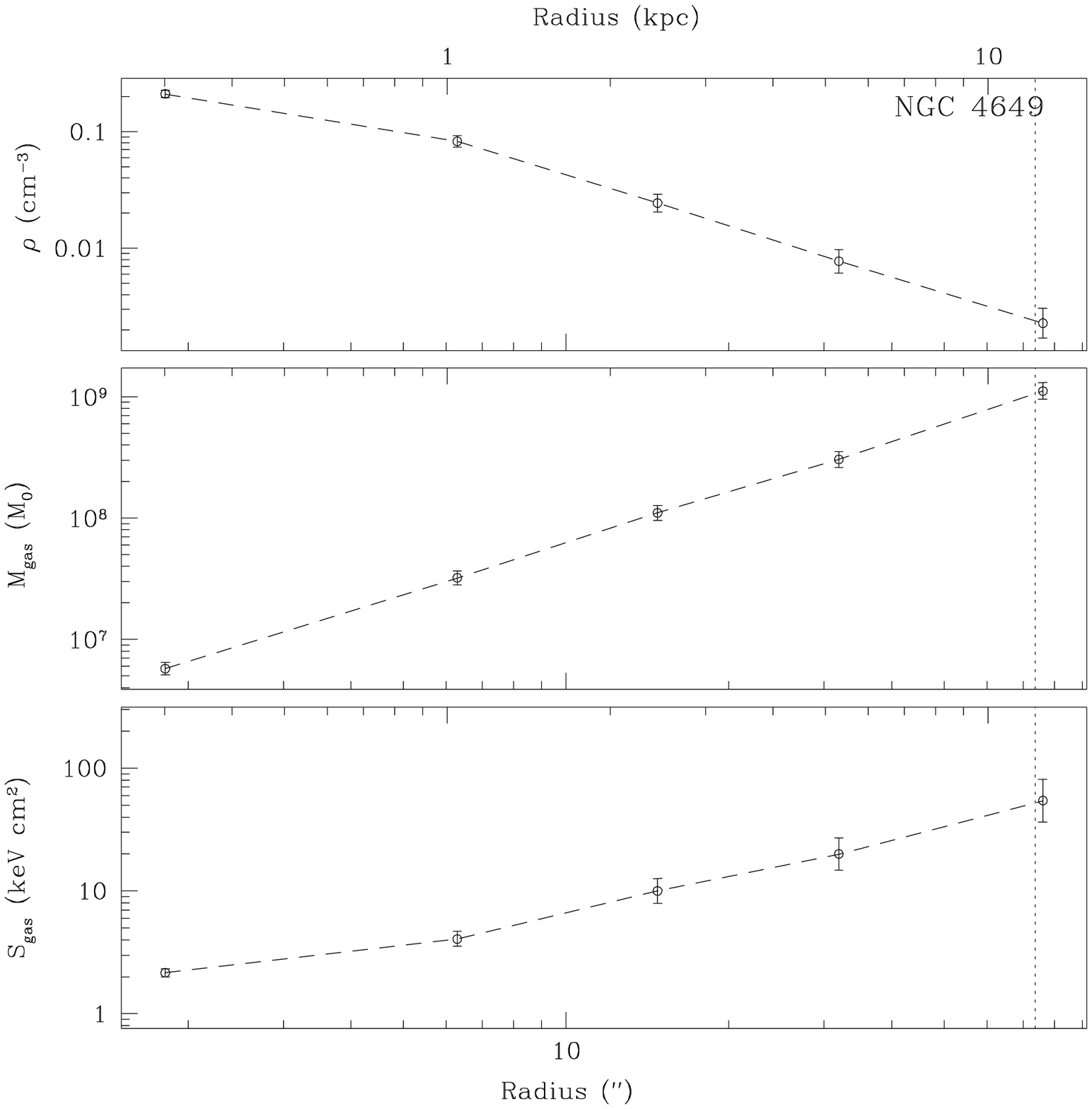}
\caption[Density, Mass, Entropy Profile: NGC 4649]
{Density, Mass and Entropy Profile for NGC 4649. Same as Figure \ref{rhomassE_NGC507}.}
\label{rhomassE_NGC4649}
\end{figure}

\begin{figure}
\epsscale{0.80}
\plotone{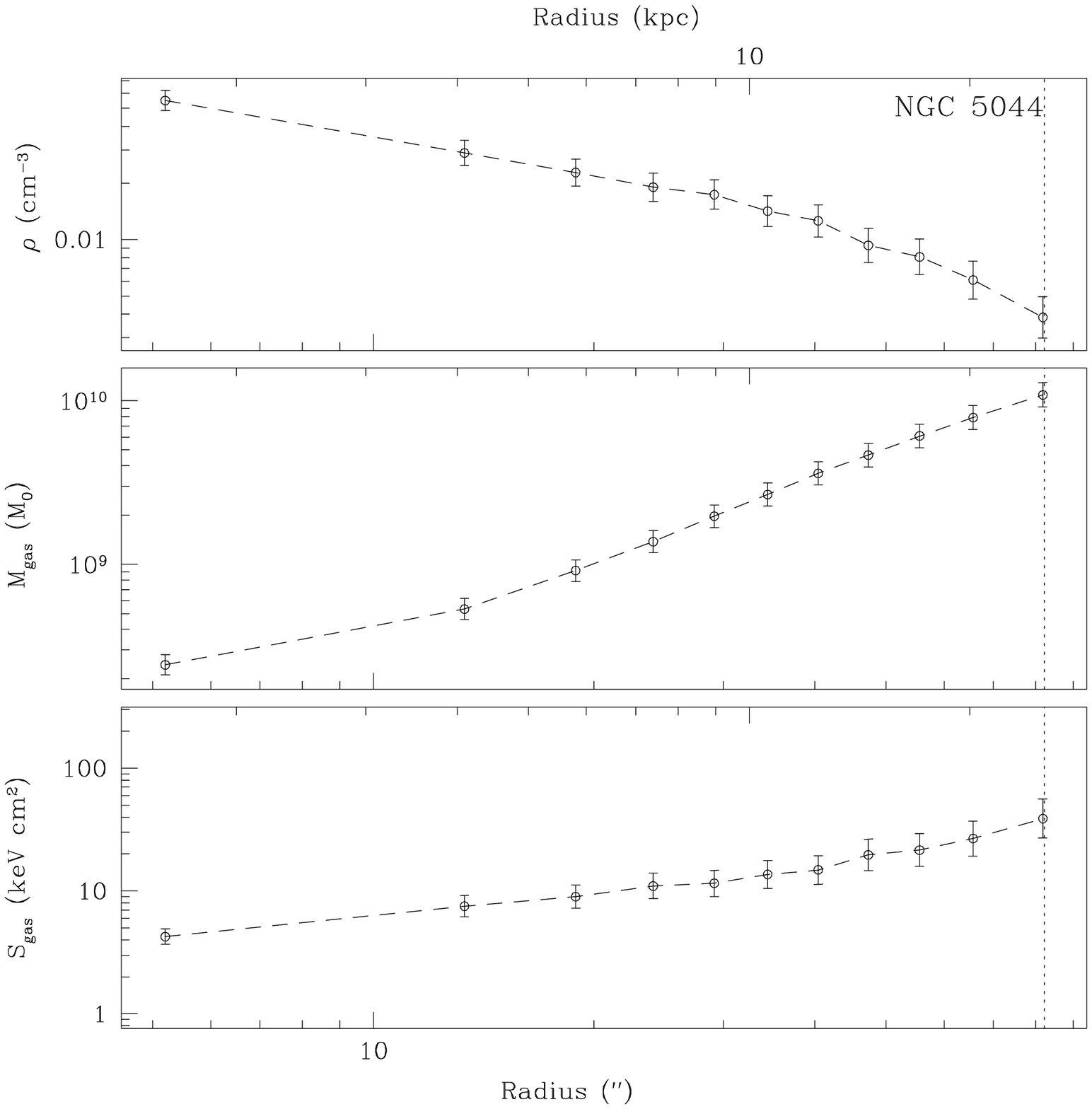}
\caption[Density, Mass, Entropy Profile: NGC 5044]
{Density, Mass and Entropy Profile for NGC 5044. Same as Figure \ref{rhomassE_NGC507}.}
\label{rhomassE_NGC5044}
\end{figure}

\begin{figure}
\epsscale{0.80}
\plotone{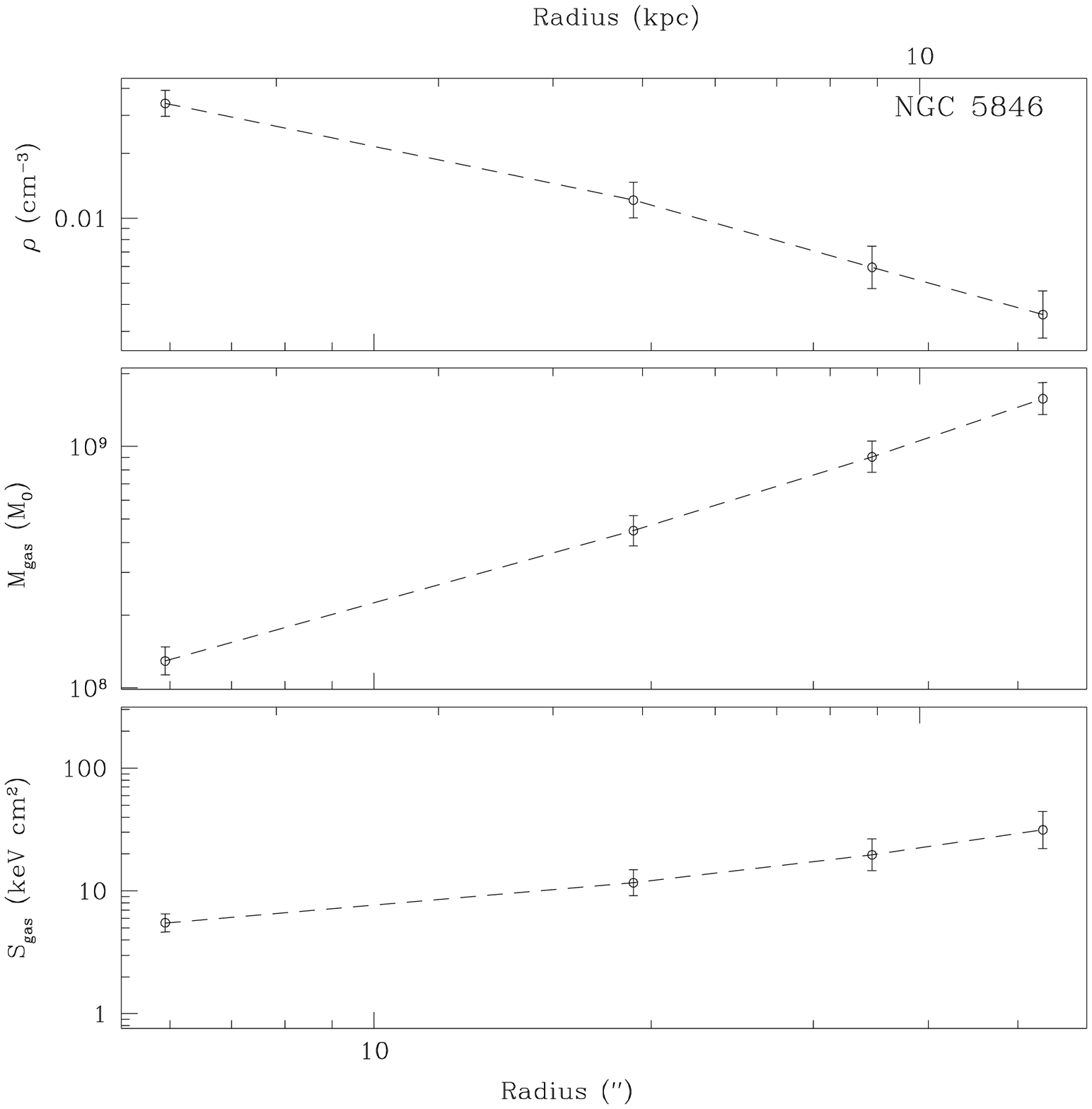}
\caption[Density, Mass, Entropy Profile: NGC 5846]
{Density, Mass and Entropy Profile for NGC 5846. Same as Figure \ref{rhomassE_NGC507}.}
\label{rhomassE_NGC5846}
\end{figure}

\begin{figure}
\epsscale{0.80}
\plotone{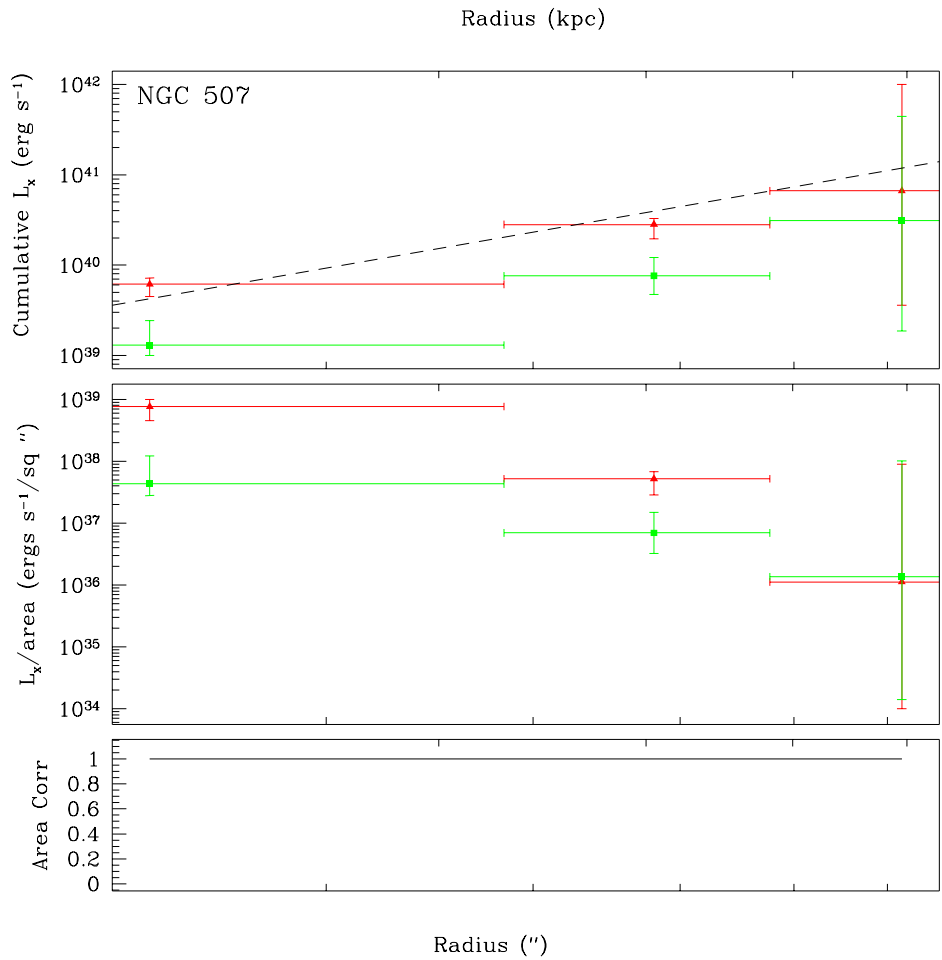}
\caption[Luminosity Profile: NGC 507]
{Luminosity Profile for NGC 507. 
All panels show radius in arseconds on the lower x-axis and radius in 
kiloparsecs on the upper x-axis.  When present, the vertical dotted line shows 
the optical half-light radius.
The top panel shows the cumulative X-ray luminosity of the gas (red triangles), the 
resolved low-mass X-ray binaries (blue circles), and the unresolved low-mass X-ray 
binaries (green squares).  The middle panel displays the same information except per area.  The 
bottom panel displays the fractional of the annular area encompassed by the ACIS-S3
observation.
}
\label{Lxc_NGC507}
\end{figure}

\begin{figure}
\epsscale{0.80}
\plotone{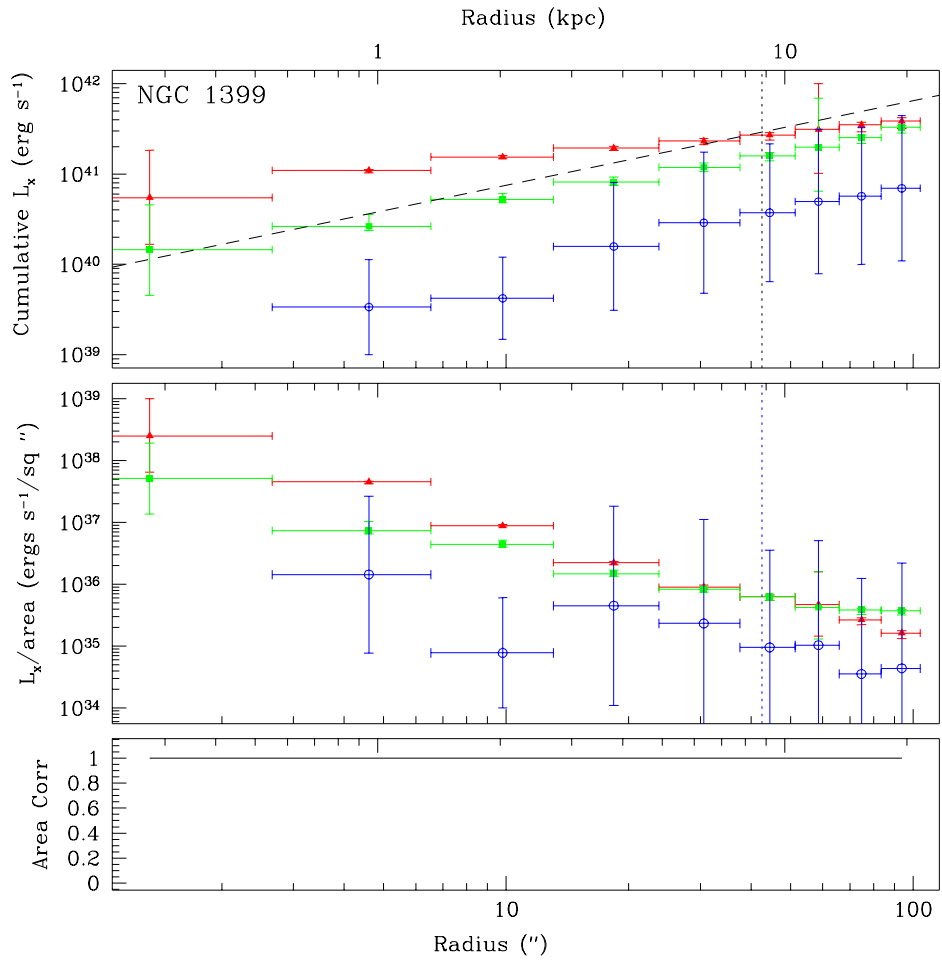}
\caption[Luminosity Profile: NGC 1399]
{Luminosity Profile for NGC 1399. Same as Figure \ref{Lxc_NGC507}.}
\label{Lxc_NGC1399}
\end{figure}

\begin{figure}
\epsscale{0.80}
\plotone{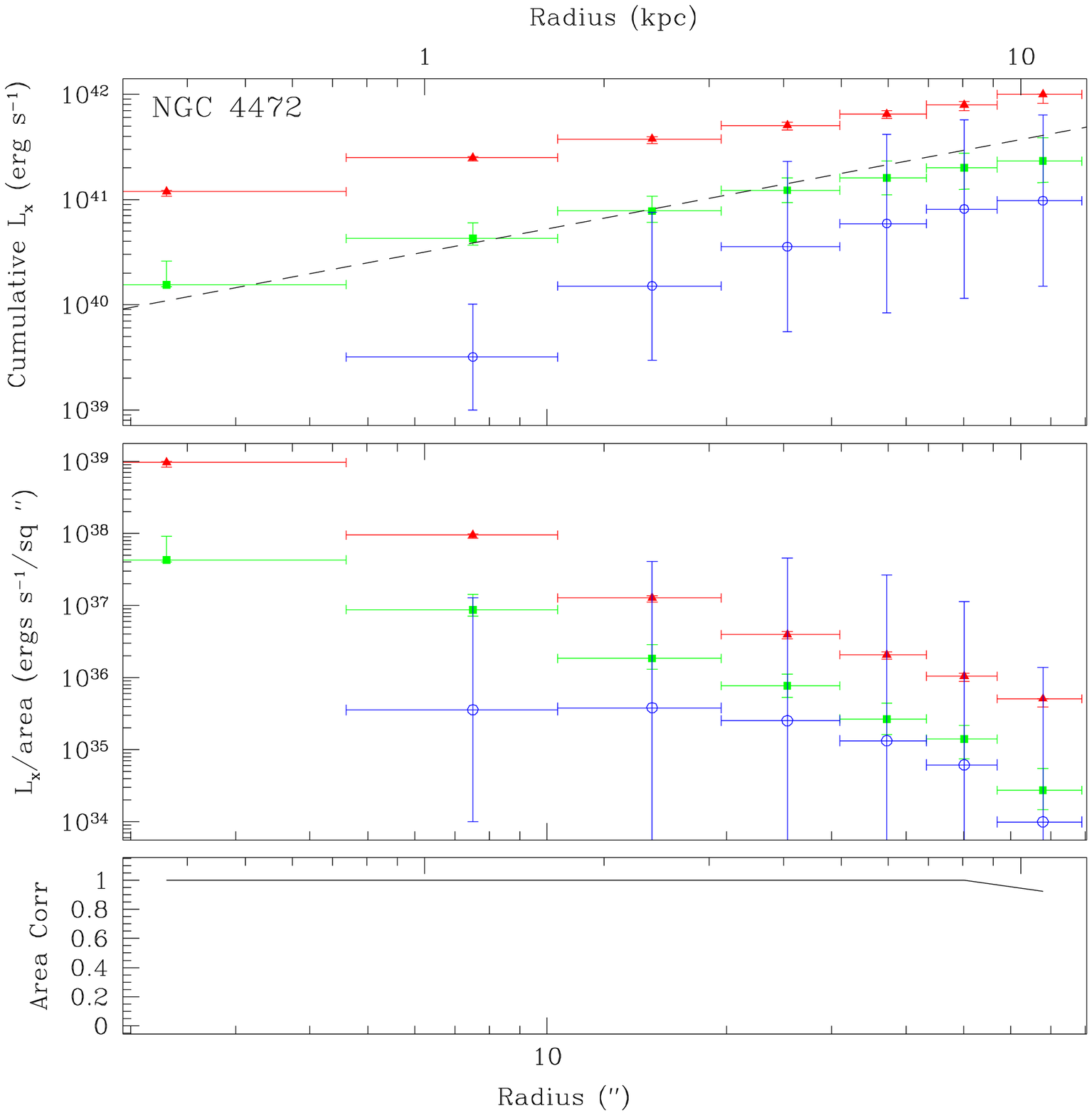}
\caption[Luminosity Profile: NGC 4472]
{Luminosity Profile for NGC 4472. Same as Figure \ref{Lxc_NGC507}.}
\label{Lxc_NGC4472}
\end{figure}

\begin{figure}
\epsscale{0.80}
\plotone{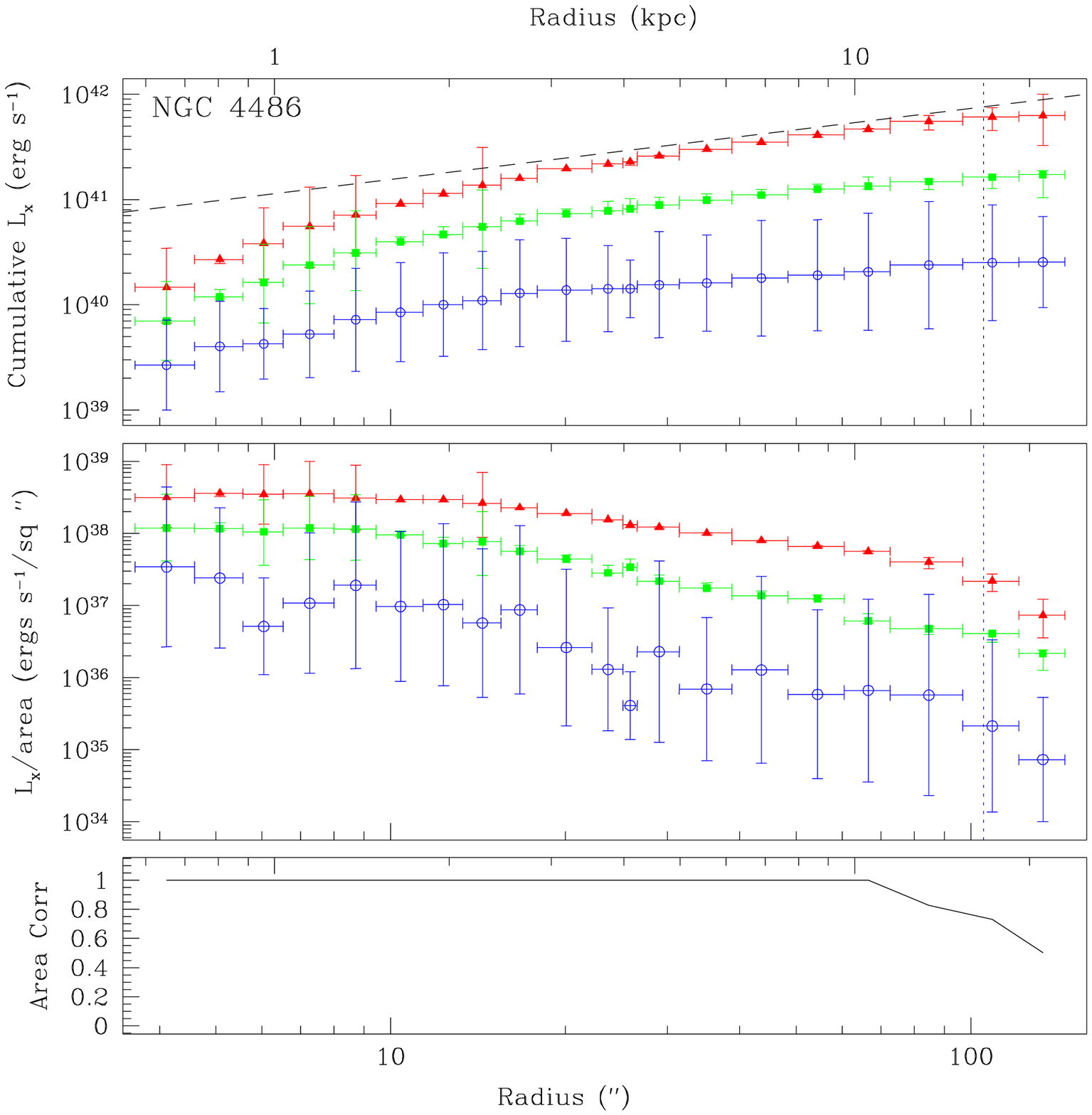}
\caption[Luminosity Profile: NGC 4486]
{Luminosity Profile for NGC 4486. Same as Figure \ref{Lxc_NGC507}.}
\label{Lxc_NGC4486}
\end{figure}

\begin{figure}
\epsscale{0.80}
\plotone{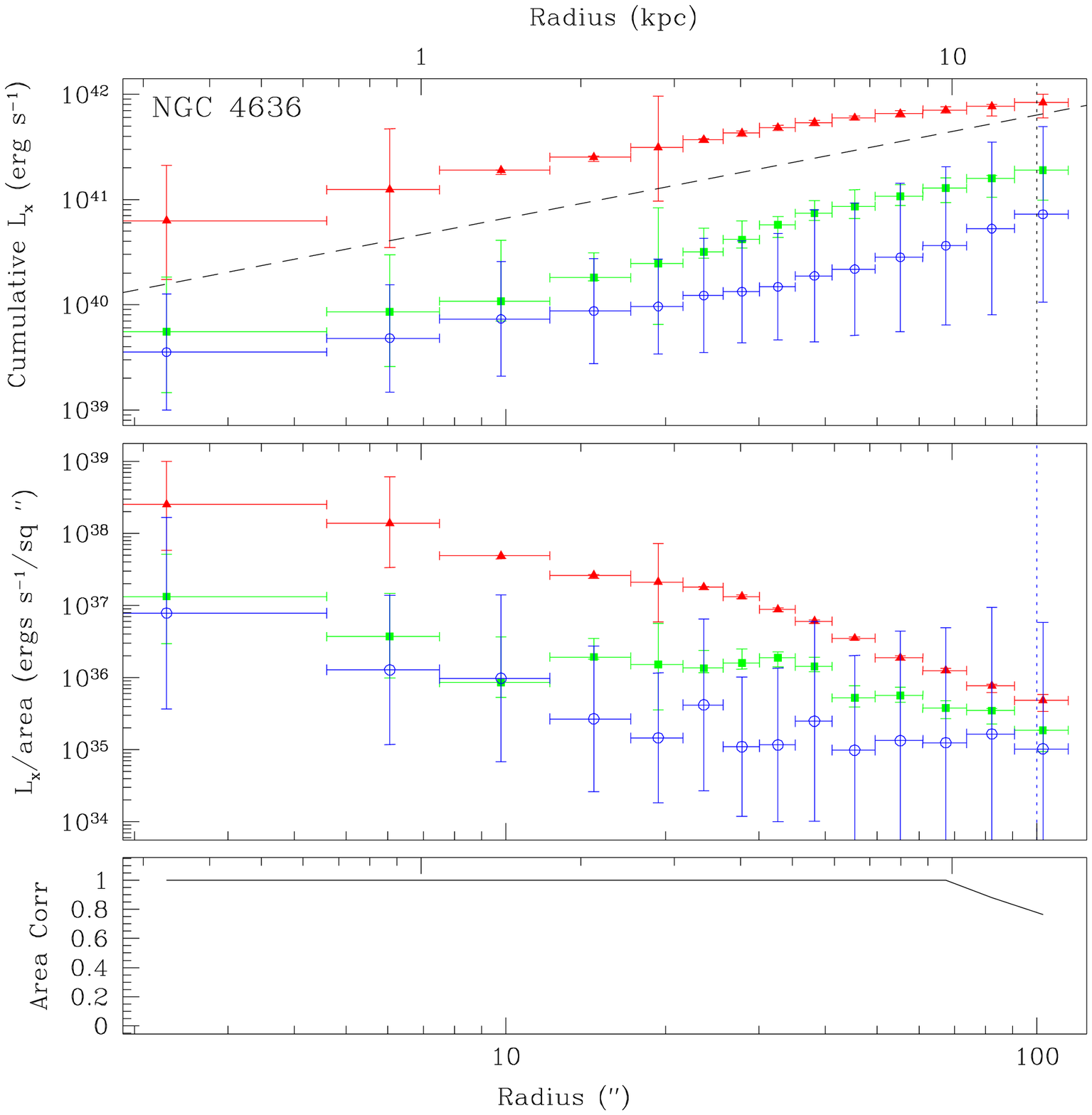}
\caption[Luminosity Profile: NGC 4636]
{Luminosity Profile for NGC 4636. Same as Figure \ref{Lxc_NGC507}.}
\label{Lxc_NGC4636}
\end{figure}

\begin{figure}
\epsscale{0.80}
\plotone{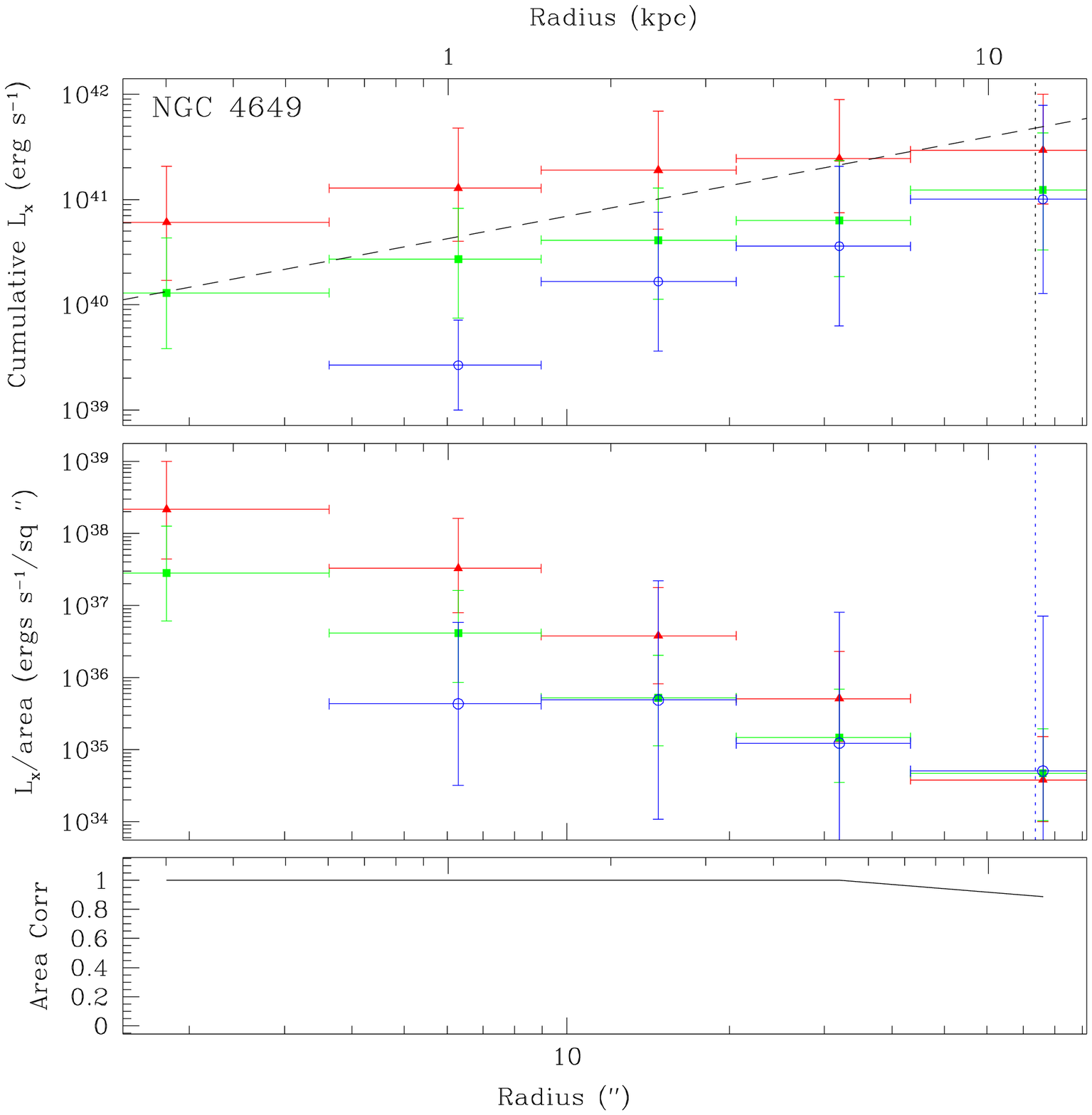}
\caption[Luminosity Profile: NGC 4649]
{Luminosity Profile for NGC 4649. Same as Figure \ref{Lxc_NGC507}.}
\label{Lxc_NGC4649}
\end{figure}

\begin{figure}
\epsscale{0.80}
\plotone{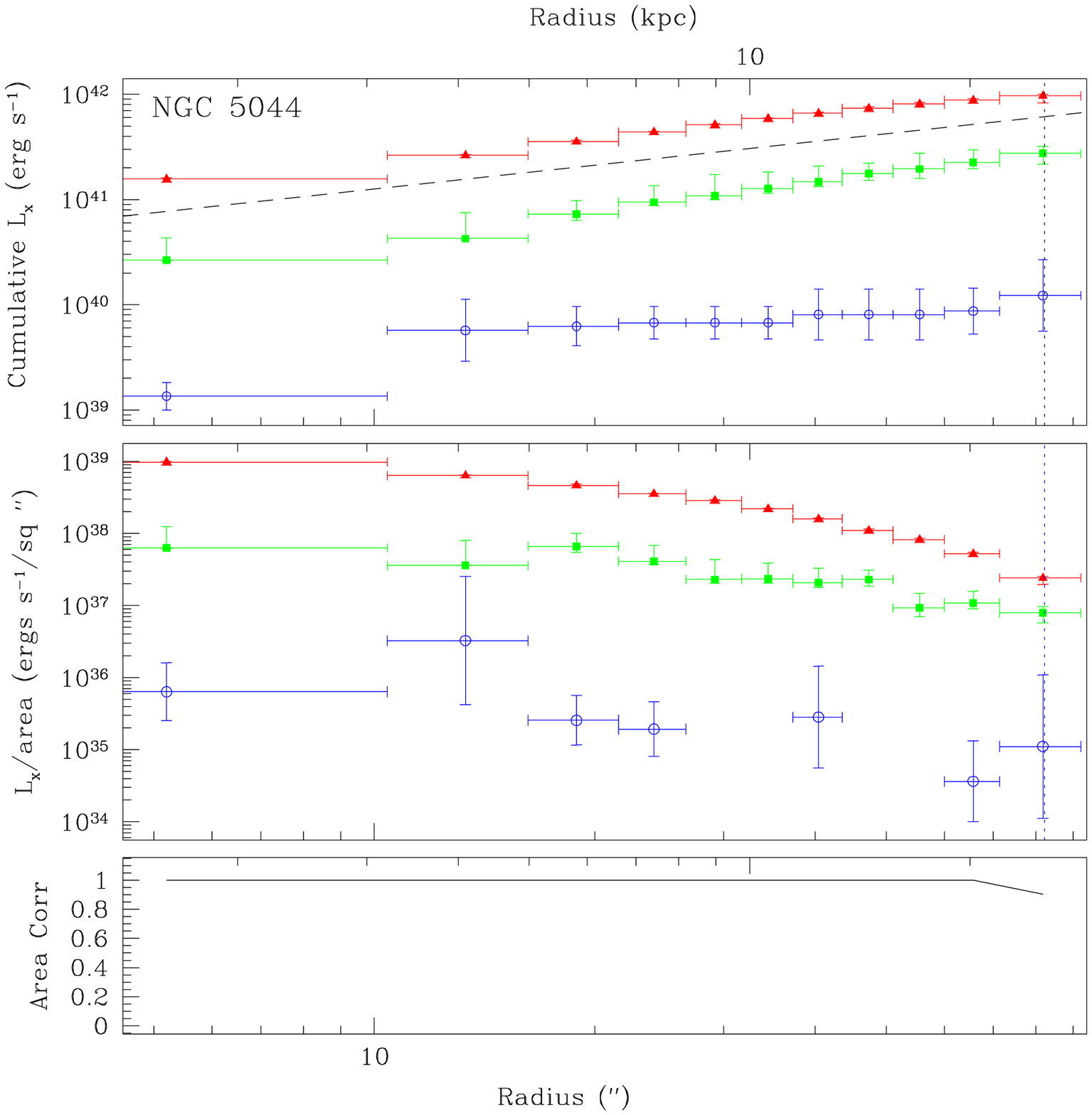}
\caption[Luminosity Profile: NGC 5044]
{Luminosity Profile for NGC 5044. Same as Figure \ref{Lxc_NGC507}.}
\label{Lxc_NGC5044}
\end{figure}

\begin{figure}
\epsscale{0.80}
\plotone{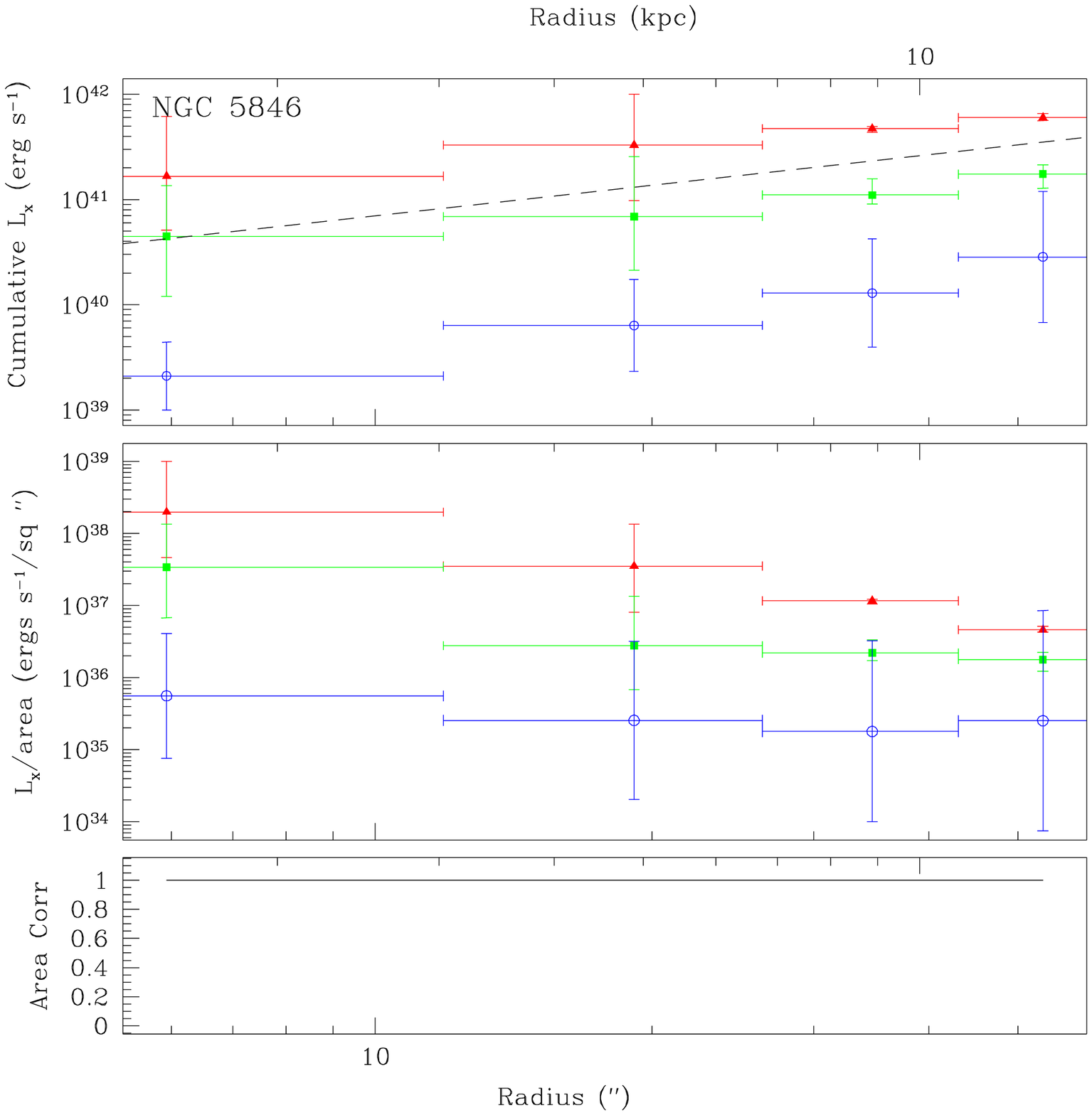}
\caption[Luminosity Profile: NGC 5846]
{Luminosity Profile for NGC 5846. Same as Figure \ref{Lxc_NGC507}.}
\label{Lxc_NGC5846}
\end{figure}

\emph{ROSAT} gas temperatures are plotted against the \emph{Chandra}
temperatures (Figure \ref{ktcomp}).  Not surprisingly, the agreement between the two measurements is poor.
This illustrates the difficulty in determining accurate plasma temperatures
when the LMXBs and central AGN cannot be removed.  If the brightest
group galaxies are selected (the filled symbols) a linear relation can 
determined, although it is not one-to-one.  These brightest group galaxies typically have much more hot gas
than LMXBs and thus the contaminations from point-sources is minimized in the 
\emph{ROSAT} determinations.

\section{Summary}

We present spectral results 
from uniformly
processed and analyzed \emph{Chandra} ACIS-S3 observations of 54 nearby
galaxies. This
sample paves the way for complete sample studies and provides a solid
basis for planning future early-type galaxy observations. In the forthcoming papers we will explore some
of the implications of this archival study.

\acknowledgements{
The author would like to acknowledge guidance from his thesis committee: Joel Bregman, Thesis advisor, Rebecca Bernstein, Mario Mateo and Timothy McKay.  Additionally, the author expresses 
gratitude for the  
useful discussions with 
Jimmy Irwin,
Renato Dupke, 
Steve Helson
John Mulchaey, 
and,
Trevor Ponmon.
}

\bibliographystyle{apj}
\bibliography{thesis,postthesis}

\begin{thebibliography}{50}
\expandafter\ifx\csname natexlab\endcsname\relax\def\natexlab#1{#1}\fi

\bibitem[{{Anders} \& {Grevesse}(1989)}]{Abundances_Anders_Grevesse}
{Anders}, E. \& {Grevesse}, N. 1989, \gca, 53, 197

\bibitem[{{Bennett} {et~al.}(2003){Bennett}, {Halpern}, {Hinshaw}, {Jarosik},
  {Kogut}, {Limon}, {Meyer}, {Page}, {Spergel}, {Tucker}, {Wollack}, {Wright},
  {Barnes}, {Greason}, {Hill}, {Komatsu}, {Nolta}, {Odegard}, {Peiris},
  {Verde}, \& {Weiland}}]{Map_preprint}
{Bennett}, C.~L., {Halpern}, M., {Hinshaw}, G., {Jarosik}, N., {Kogut}, A.,
  {Limon}, M., {Meyer}, S.~S., {Page}, L., {Spergel}, D.~N., {Tucker}, G.~S.,
  {Wollack}, E., {Wright}, E.~L., {Barnes}, C., {Greason}, M.~R., {Hill},
  R.~S., {Komatsu}, E., {Nolta}, M.~R., {Odegard}, N., {Peiris}, H.~V.,
  {Verde}, L., \& {Weiland}, J.~L. 2003, \apjs, 148, 1

\bibitem[{{Beuing} {et~al.}(1999){Beuing}, {Dobereiner}, {Bohringer}, \&
  {Bender}}]{ROSAT_RASS_survey}
{Beuing}, J., {Dobereiner}, S., {Bohringer}, H., \& {Bender}, R. 1999, \mnras,
  302, 209

\bibitem[{{Bregman} {et~al.}(1995){Bregman}, {Hogg}, \&
  {Roberts}}]{Lx_ROSAT_NGC1291}
{Bregman}, J.~N., {Hogg}, D.~E., \& {Roberts}, M.~S. 1995, \apj, 441, 561

\bibitem[{{Bregman} {et~al.}(2001){Bregman}, {Miller}, \&
  {Irwin}}]{Joel_OVI_FUSE}
{Bregman}, J.~N., {Miller}, E.~D., \& {Irwin}, J.~A. 2001, \apjl, 553, L125

\bibitem[{{Brown} \& {Bregman}(1998)}]{Brown_Bregman}
{Brown}, B.~A. \& {Bregman}, J.~N. 1998, \apjl, 495, L75

\bibitem[{{Buote}(1999)}]{Buote_2-Temps_1}
{Buote}, D.~A. 1999, \mnras, 309, 685

\bibitem[{{Buote} \& {Fabian}(1998)}]{Buote_2-Temps_2}
{Buote}, D.~A. \& {Fabian}, A.~C. 1998, \mnras, 296, 977

\bibitem[{{Burstein} {et~al.}(1987){Burstein}, {Davies}, {Dressler}, {Faber},
  {Stone}, {Lynden-Bell}, {Terlevich}, \& {Wegner}}]{B87}
{Burstein}, D., {Davies}, R.~L., {Dressler}, A., {Faber}, S.~M., {Stone},
  R.~P.~S., {Lynden-Bell}, D., {Terlevich}, R.~J., \& {Wegner}, G. 1987, \apjs,
  64, 601

\bibitem[{{Canizares} {et~al.}(1987){Canizares}, {Fabbiano}, \&
  {Trinchieri}}]{LxLb_1}
{Canizares}, C.~R., {Fabbiano}, G., \& {Trinchieri}, G. 1987, \apj, 312, 503

\bibitem[{{Capaccioli} {et~al.}(1987){Capaccioli}, {Held}, \& {Nieto}}]{C87}
{Capaccioli}, M., {Held}, E.~V., \& {Nieto}, J. 1987, \aj, 94, 1519

\bibitem[{{Colbert} {et~al.}(2001){Colbert}, {Mulchaey}, \&
  {Zabludoff}}]{Isolated_ETGs_Mulchaey}
{Colbert}, J.~W., {Mulchaey}, J.~S., \& {Zabludoff}, A.~I. 2001, \aj, 121, 808

\bibitem[{{Davis} \& {White}(1996)}]{Davis_White_ROSAT}
{Davis}, D.~S. \& {White}, R.~E. 1996, \apjl, 470, L35

\bibitem[{{de Vaucouleurs} {et~al.}(1991){de Vaucouleurs}, {de Vaucouleurs},
  {Corwin}, {Buta}, {Paturel}, \& {Fouque}}]{RC3}
{de Vaucouleurs}, G., {de Vaucouleurs}, A., {Corwin}, H.~G., {Buta}, R.~J.,
  {Paturel}, G., \& {Fouque}, P. 1991, {Third Reference Catalogue of Bright
  Galaxies} (Volume 1-3, XII, 2069 pp.~7 figs..~ Springer-Verlag Berlin
  Heidelberg New York)

\bibitem[{{Dickey} \& {Lockman}(1990)}]{nH_Dickey_Lockman}
{Dickey}, J.~M. \& {Lockman}, F.~J. 1990, \araa, 28, 215

\bibitem[{{Faber} {et~al.}(1989){Faber}, {Wegner}, {Burstein}, {Davies},
  {Dressler}, {Lynden-Bell}, \& {Terlevich}}]{Faber_89}
{Faber}, S.~M., {Wegner}, G., {Burstein}, D., {Davies}, R.~L., {Dressler}, A.,
  {Lynden-Bell}, D., \& {Terlevich}, R.~J. 1989, \apjs, 69, 763

\bibitem[{{Forman} {et~al.}(1985){Forman}, {Jones}, \& {Tucker}}]{Einstein_2}
{Forman}, W., {Jones}, C., \& {Tucker}, W. 1985, \apj, 293, 102

\bibitem[{{Forman} {et~al.}(1979){Forman}, {Schwarz}, {Jones}, {Liller}, \&
  {Fabian}}]{Forman_Einstein}
{Forman}, W., {Schwarz}, J., {Jones}, C., {Liller}, W., \& {Fabian}, A.~C.
  1979, \apjl, 234, L27

\bibitem[{{Franx} {et~al.}(1989){Franx}, {Illingworth}, \& {Heckman}}]{F89}
{Franx}, M., {Illingworth}, G., \& {Heckman}, T. 1989, \aj, 98, 538

\bibitem[{{Freedman} {et~al.}(2001){Freedman}, {Madore}, {Gibson}, {Ferrarese},
  {Kelson}, {Sakai}, {Mould}, {Kennicutt}, {Ford}, {Graham}, {Huchra},
  {Hughes}, {Illingworth}, {Macri}, \& {Stetson}}]{HST_Key_Ho}
{Freedman}, W.~L., {Madore}, B.~F., {Gibson}, B.~K., {Ferrarese}, L., {Kelson},
  D.~D., {Sakai}, S., {Mould}, J.~R., {Kennicutt}, R.~C., {Ford}, H.~C.,
  {Graham}, J.~A., {Huchra}, J.~P., {Hughes}, S.~M.~G., {Illingworth}, G.~D.,
  {Macri}, L.~M., \& {Stetson}, P.~B. 2001, \apj, 553, 47

\bibitem[{{Gibson} {et~al.}(2000){Gibson}, {Stetson}, {Freedman}, {Mould},
  {Kennicutt}, {Huchra}, {Sakai}, {Graham}, {Fassett}, {Kelson}, {Ferrarese},
  {Hughes}, {Illingworth}, {Macri}, {Madore}, {Sebo}, \&
  {Silbermann}}]{dist_NGC5253}
{Gibson}, B.~K., {Stetson}, P.~B., {Freedman}, W.~L., {Mould}, J.~R.,
  {Kennicutt}, R.~C., {Huchra}, J.~P., {Sakai}, S., {Graham}, J.~A., {Fassett},
  C.~I., {Kelson}, D.~D., {Ferrarese}, L., {Hughes}, S.~M.~G., {Illingworth},
  G.~D., {Macri}, L.~M., {Madore}, B.~F., {Sebo}, K.~M., \& {Silbermann}, N.~A.
  2000, \apj, 529, 723

\bibitem[{{Goudfrooij} {et~al.}(1994{\natexlab{a}}){Goudfrooij}, {Hansen},
  {Jorgensen}, \& {Norgaard-Nielsen}}]{Goudfrooij_2}
{Goudfrooij}, P., {Hansen}, L., {Jorgensen}, H.~E., \& {Norgaard-Nielsen},
  H.~U. 1994{\natexlab{a}}, \aaps, 105, 341

\bibitem[{{Goudfrooij} {et~al.}(1994{\natexlab{b}}){Goudfrooij}, {Hansen},
  {Jorgensen}, {Norgaard-Nielsen}, {de Jong}, \& {van den Hoek}}]{Goudfrooij_1}
{Goudfrooij}, P., {Hansen}, L., {Jorgensen}, H.~E., {Norgaard-Nielsen}, H.~U.,
  {de Jong}, T., \& {van den Hoek}, L.~B. 1994{\natexlab{b}}, \aaps, 104, 179

\bibitem[{{Helsdon} {et~al.}(2001){Helsdon}, {Ponman}, {O'Sullivan}, \&
  {Forbes}}]{Helsdon_1}
{Helsdon}, S.~F., {Ponman}, T.~J., {O'Sullivan}, E., \& {Forbes}, D.~A. 2001,
  \mnras, 325, 693

\bibitem[{{Irwin} {et~al.}(2003){Irwin}, {Athey}, \& {Bregman}}]{Jimmy_LMXB}
{Irwin}, J.~A., {Athey}, A.~E., \& {Bregman}, J.~N. 2003, \apj, 587, 356

\bibitem[{{Irwin} {et~al.}(2004){Irwin}, {Bregman}, \& {Athey}}]{Jimmy_ULX}
{Irwin}, J.~A., {Bregman}, J.~N., \& {Athey}, A.~E. 2004, \apjl, 601, L143

\bibitem[{{Irwin} \& {Sarazin}(1998)}]{Irwin_Sarazin_ROSAT}
{Irwin}, J.~A. \& {Sarazin}, C.~L. 1998, \apj, 499, 650

\bibitem[{{Jorgensen} {et~al.}(1995){Jorgensen}, {Franx}, \&
  {Kjaergaard}}]{J95}
{Jorgensen}, I., {Franx}, M., \& {Kjaergaard}, P. 1995, \mnras, 273, 1097

\bibitem[{{Macchetto} {et~al.}(1996){Macchetto}, {Pastoriza}, {Caon}, {Sparks},
  {Giavalisco}, {Bender}, \& {Capaccioli}}]{Caon_paper_I}
{Macchetto}, F., {Pastoriza}, M., {Caon}, N., {Sparks}, W.~B., {Giavalisco},
  M., {Bender}, R., \& {Capaccioli}, M. 1996, \aaps, 120, 463

\bibitem[{{Markevitch}(2002)}]{Markevitch}
{Markevitch}, M. 2002, astro-ph/0205333

\bibitem[{{Markevitch} {et~al.}(2003){Markevitch}, {Bautz}, {Biller}, {Butt},
  {Edgar}, {Gaetz}, {Garmire}, {Grant}, {Green}, {Juda}, {Plucinsky},
  {Schwartz}, {Smith}, {Vikhlinin}, {Virani}, {Wargelin}, \&
  {Wolk}}]{Chandra_Absorption}
{Markevitch}, M., {Bautz}, M.~W., {Biller}, B., {Butt}, Y., {Edgar}, R.,
  {Gaetz}, T., {Garmire}, G., {Grant}, C.~E., {Green}, P., {Juda}, M.,
  {Plucinsky}, P.~P., {Schwartz}, D., {Smith}, R., {Vikhlinin}, A., {Virani},
  S., {Wargelin}, B.~J., \& {Wolk}, S. 2003, \apj, 583, 70

\bibitem[{{Matsushita}(2001)}]{ROSAT_Large_Study_Matsushita}
{Matsushita}, K. 2001, \apj, 547, 693

\bibitem[{{Matsushita} {et~al.}(1997){Matsushita}, {Makishima}, {Rokutanda},
  {Yamasaki}, \& {Ohashi}}]{ASCA_NGC4636}
{Matsushita}, K., {Makishima}, K., {Rokutanda}, E., {Yamasaki}, N.~Y., \&
  {Ohashi}, T. 1997, \apjl, 488, L125

\bibitem[{{Matsushita} {et~al.}(2000){Matsushita}, {Ohashi}, \&
  {Makishima}}]{ASCA_27_galaxies}
{Matsushita}, K., {Ohashi}, T., \& {Makishima}, K. 2000, \pasj, 52, 685

\bibitem[{{Mei} {et~al.}(2000){Mei}, {Silva}, \& {Quinn}}]{dist_IC4296}
{Mei}, S., {Silva}, D., \& {Quinn}, P.~J. 2000, \aap, 361, 68

\bibitem[{{Mulchaey} \& {Zabludoff}(1999)}]{Lx_ROSAT_NGC1132}
{Mulchaey}, J.~S. \& {Zabludoff}, A.~I. 1999, \apj, 514, 133

\bibitem[{{Mushotzky} {et~al.}(1994){Mushotzky}, {Loewenstein}, {Awaki},
  {Makishima}, {Matsushita}, \& {Matsumoto}}]{ASCA_NGC4636_Mushotzky}
{Mushotzky}, R.~F., {Loewenstein}, M., {Awaki}, H., {Makishima}, K.,
  {Matsushita}, K., \& {Matsumoto}, H. 1994, \apjl, 436, L79

\bibitem[{{O'Sullivan} {et~al.}(2001){O'Sullivan}, {Forbes}, \&
  {Ponman}}]{Trevor_ROSAT_study}
{O'Sullivan}, E., {Forbes}, D.~A., \& {Ponman}, T.~J. 2001, \mnras, 328, 461

\bibitem[{{O'Sullivan} {et~al.}(2003){O'Sullivan}, {Ponman}, \&
  {Collins}}]{Trevor_ROSAT_scaling}
{O'Sullivan}, E., {Ponman}, T.~J., \& {Collins}, R.~S. 2003, \mnras, 340, 1375

\bibitem[{{Peletier} {et~al.}(1990){Peletier}, {Davies}, {Illingworth},
  {Davis}, \& {Cawson}}]{P90}
{Peletier}, R.~F., {Davies}, R.~L., {Illingworth}, G.~D., {Davis}, L.~E., \&
  {Cawson}, M. 1990, \aj, 100, 1091

\bibitem[{{Ryden} {et~al.}(2001){Ryden}, {Forbes}, \& {Terlevich}}]{R01}
{Ryden}, B.~S., {Forbes}, D.~A., \& {Terlevich}, A.~I. 2001, \mnras, 326, 1141

\bibitem[{{Sarazin} \& {Bahcall}(1977)}]{Beta_model_Sarazin77}
{Sarazin}, C.~L. \& {Bahcall}, J.~N. 1977, \apjs, 34, 451

\bibitem[{{Smith} {et~al.}(2001){Smith}, {Brickhouse}, {Liedahl}, \&
  {Raymond}}]{APEC}
{Smith}, R.~K., {Brickhouse}, N.~S., {Liedahl}, D.~A., \& {Raymond}, J.~C.
  2001, \apjl, 556, L91

\bibitem[{{Tamura} {et~al.}(2003){Tamura}, {Kaastra}, {Makishima}, \&
  {Takahashi}}]{XMMRGS_NGC5044}
{Tamura}, T., {Kaastra}, J.~S., {Makishima}, K., \& {Takahashi}, I. 2003, \aap,
  399, 497

\bibitem[{{Terlevich} \& {Forbes}(2002)}]{Forbes_Ages}
{Terlevich}, A.~I. \& {Forbes}, D.~A. 2002, \mnras, 330, 547

\bibitem[{{Tonry} {et~al.}(2001){Tonry}, {Dressler}, {Blakeslee}, {Ajhar},
  {Fletcher}, {Luppino}, {Metzger}, \& {Moore}}]{Tonry_SBF_distances}
{Tonry}, J.~L., {Dressler}, A., {Blakeslee}, J.~P., {Ajhar}, E.~A., {Fletcher},
  A.~B., {Luppino}, G.~A., {Metzger}, M.~R., \& {Moore}, C.~B. 2001, \apj, 546,
  681

\bibitem[{{Toomre}(1977)}]{NGC7252_is_strange_2}
{Toomre}, A. 1977, \araa, 15, 437

\bibitem[{{Trager} {et~al.}(2000{\natexlab{a}}){Trager}, {Faber}, {Worthey}, \&
  {Gonz{\' a}lez}}]{Trager_2}
{Trager}, S.~C., {Faber}, S.~M., {Worthey}, G., \& {Gonz{\' a}lez}, J.~J.
  2000{\natexlab{a}}, \aj, 120, 165

\bibitem[{{Trager} {et~al.}(2000{\natexlab{b}}){Trager}, {Faber}, {Worthey}, \&
  {Gonz{\' a}lez}}]{Trager_1}
---. 2000{\natexlab{b}}, \aj, 119, 1645

\bibitem[{{Xu} {et~al.}(2002){Xu}, {Kahn}, {Peterson}, {Behar}, {Paerels},
  {Mushotzky}, {Jernigan}, {Brinkman}, \& {Makishima}}]{XMMRGS_NGC4636}
{Xu}, H., {Kahn}, S.~M., {Peterson}, J.~R., {Behar}, E., {Paerels}, F.~B.~S.,
  {Mushotzky}, R.~F., {Jernigan}, J.~G., {Brinkman}, A.~C., \& {Makishima}, K.
  2002, \apj, 579, 600

\end{thebibliography}

\onecolumn

\clearpage

\clearpage

\clearpage

\clearpage

\clearpage

\clearpage

\clearpage

\clearpage

\clearpage

\clearpage

\clearpage

\clearpage

\clearpage

\clearpage

\clearpage

\clearpage

\clearpage

\clearpage

\clearpage

\clearpage

\clearpage

\clearpage

\clearpage

\clearpage

\clearpage

\clearpage

\clearpage

\clearpage

\clearpage





\end{document}